\documentclass[]{interact}

\usepackage{epstopdf}
\usepackage{xcolor}
\usepackage{bigints}
\usepackage{amsmath}
\usepackage{subfigure}
\usepackage{tikz}
\usepackage{braket}

\usepackage[numbers,sort&compress]{natbib}
\bibpunct[, ]{[}{]}{,}{n}{,}{,}
\makeatletter
\def\NAT@def@citea{\def\@citea{\NAT@separator}}
\makeatother

\theoremstyle{plain}

\theoremstyle{definition}

\theoremstyle{remark}

\newcommand*\circled[1]{\tikz[baseline=(char.base)]{
            \node[shape=circle,draw,inner sep=2pt] (char) {#1};}}

\begin{document}


\title{Astrophotonics: recent and future developments}

\author{
\name{Simon Charles Ellis\textsuperscript{a}\thanks{CONTACT Simon Ellis. Email: simon.ellis@mq.edu.au} and Joss Bland-Hawthorn\textsuperscript{b}}
\affil{\textsuperscript{a}Australian Astronomical Optics, Macquarie University, Australia;\\
\textsuperscript{b}Sydney Institute for Astronomy, School of Physics A28, University of Sydney, Australia}
}

\maketitle

\begin{abstract}
Astrophotonics is a burgeoning field that lies at the interface of photonics and modern astronomical instrumentation. Here we provide a pedagogical review of basic photonic functions that enable modern instruments, and give an overview of recent and future applications.
Traditionally, optical fibres have been used in innovative ways to vastly increase the multiplex advantage of an astronomical instrument, e.g.\ the ability to observe hundreds or thousands of stars simultaneously. But modern instruments are using many new photonic functions, some emerging from the telecom industry, and others specific to the demands of adaptive optics systems on modern telescopes.    As telescopes continue to increase in size, we look to a future where instruments exploit the properties of individual photons. In particular, we envisage telescopes and interferometers that build on international
developments in quantum networks, the so-called quantum internet. With the aid of entangled photons and quantum logic gates, the new infrastructures seek to preserve the photonic state and timing of individual photons over a coherent network.
\end{abstract}


\begin{keywords}
Sections; lists; figures; tables; mathematics; fonts; references; appendices
\end{keywords}

\section{Introduction}

Modern photonics, as distinct from optics, provides us with revolutionary, new ways to ``mould the flow of light'' through an apparatus. In practice, modern optical systems make increasing use of photonics, and vice versa, such that the boundaries between what we mean by optical and photonic instruments are becoming increasingly blurred. 
Nevertheless, the practical application of photonics to optical astronomical instrumentation, i.e., astrophotonics, is a significant shift from traditional approaches to astronomical instrumentation.  Whereas classical instruments working at visible and near-infrared wavelengths (i.e.\ approximately 350 -- 5000~nm) have used bulk optics. i.e.\ lenses, mirrors, prisms, gratings etc., to manipulate beams of light propagating through free-space, astrophotonics employs photonic devices to manipulate the light within optical waveguides.  Many photonic functions have equivalent optical transforms, e.g.\ the use of a fibre taper to change a beam's properties is similar to the action of a focal enlarger/reducer in an astronomical instrument. But some functions, such as the conversion from multimode to single mode behavious in a photonic lantern, have 
no optical equivalent.

The photonic revolution has been driven by the demands of modern telecom, and made possible by vast improvements in the purity and malleability of materials. In particular, drawing towers can produce a silica fibre that has insertion losses of less than 0.1~dB over a length of 1~km. Over that same length, the best modern glasses would be opaque.

Astrophotonics is a rapidly developing field,  as evidenced by the breadth of research presented in previous reviews\cite{bland12,bland17b,gat19,min21};
special journal issues\cite{bland09,bry17,din21a,din21b}; 
an up-to-date and greatly expanded discussion is provided in a series of white papers\cite{jov23}.
Our recent textbook provides a pedagogical introduction to the field\cite{ell23}.  
Moreover, astrophotonics is transitioning from an active field of research and development to a mainstream branch of astronomical instrumentation.  Photonic devices and components are now being used in facility instruments, e.g.\ ESO's GRAVITY\cite{grav17}, and many others are being demonstrated in on-sky experiments.
Here we present a brief summary of the latest developments in astrophotonics, and look to future developments aimed at exploiting the full potential of the next generation of extremely large telescopes, and the properties of individual photons. 


 


\section{Photonic functions}
\label{sec:photonicfns}

We begin by describing  fundamental photonic functions and the devices that accomplish these.  These devices are the building-blocks for more complicated photonic circuits and instruments.  The basic mechanisms described here will be expanded upon in the practical applications discussed in section~\ref{sec:pracapp}.














\subsection{Transporting light}
\label{sec:transport}

By far the commonest
application of photonics in astronomy is the use of fibre optics to transport light from one part of the optical train to another.   The pre-eminent example of this is found in multi-object spectroscopy (MOS), in which light is transported from the focal plane of the telescope to the pseudo entrance  slit of a spectrometer, allowing highly multiplexed observations (100s or 1000s of objects), access to the full telescope field-of-view and homogeneous wavelength coverage.  Indeed, this was the first application of astrophotonics\cite{hil80} and subsequently has been responsible for profound advances in our understanding of cosmology (notably structure formation and evolution e.g.\cite{col01,yor00}), galaxy evolution\cite{dri11}, and the formation of our own Galaxy\cite{des15}; the impact of which is hard to overstate.

Although the scientific impact of this use of fibres has been enormous, from a functional viewpoint this is the simplest application of photonics: light injected into one face of a fibre (or other waveguide) is transported to the other end of the fibre.
For typical step-index fibres this is a result of total internal reflection at the core-cladding interface.  Since this can only occur for light propagating along the optical axis at angles less than the critical angle for total internal reflection,
this leads to a maximum injection angle, $\phi$, given by
\begin{equation}
\label{e:na}
    n_{0} \sin \phi \le \sqrt{n_{\rm core}^{2} - n_{\rm clad}^{2}},
\end{equation}
where the quantity $n_{0} \sin \phi$ is called the numerical aperture (NA) of the fibre.


An important principle for any optical system, including optical fibres and waveguides, is the conservation of \'{e}tendue (sometimes known as the A$\Omega$ product in astronomy), which states that the \'{e}tendue, 
\begin{equation}
\epsilon = n^{2} A \Omega,
\end{equation}
    cannot be decreased in an optical system without loss of light, and $n$ is the refractive index of the medium, $A$ is the area of the beam, and $\Omega$ is the solid angle.  For waveguides, the \'{e}tendue is related to the numerical aperture, such that
    \begin{equation}
        \epsilon = \pi A N\!\!A^{2}.
    \end{equation}

For optical fibres an important caveat to the conservation of \'{e}tendue occurs when a fibre is injected with a beam that is substantially slower (i.e.\ at a higher focal ratio, or smaller NA) than the NA of the fibre itself.  The emerging beam will have an faster focal ratio (higher NA) than the input, a phenomenon known as focal ratio degradation (FRD) or NA up-scattering.  FRD thus increases the \'{e}tendue of the system, and all subsequent optics must therefore be either faster or larger as a consequence, and even then will lead to a loss of resolving power, multilplex or wavelength coverage, or else will lead to a loss of throughput if the FRD is not compensated for.

\begin{figure}
    \centering
    \includegraphics[scale=0.45]{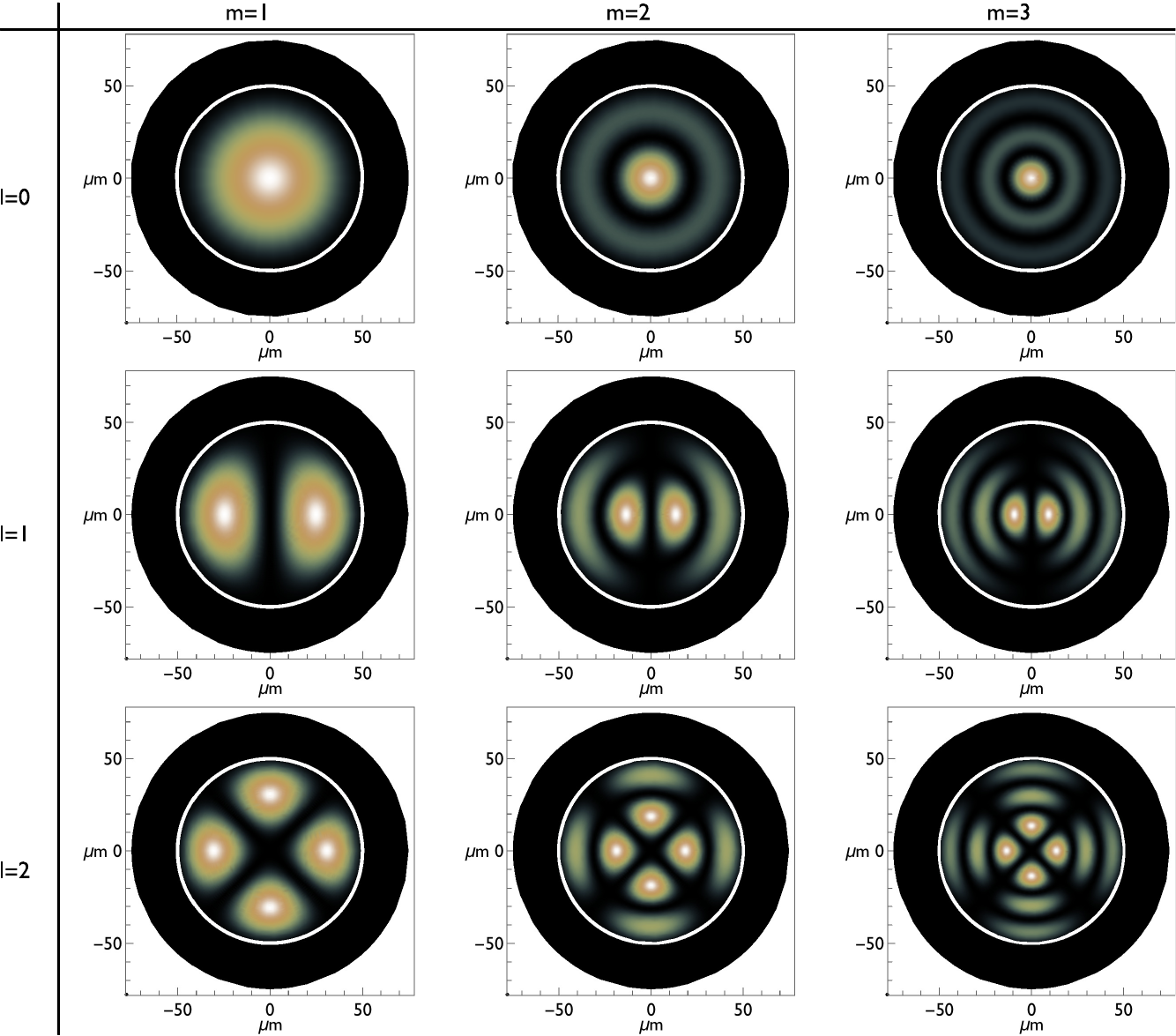}
    \caption{The electric field distribution for the first 9 modes of a fibre with $a=50$~$\mu$m, $n_{\rm core}=1.51$ and $n_{\rm clad}=1.52$ at $\lambda_{0} = 1.55$~$\mu$m.}
    \label{f:fibremodes}
\end{figure}

An important concept for the understanding and application of photonics is that of waveguide modes.  Equation~\ref{e:na} defines the maximum on-axis injection angle into a step-index waveguide, but light cannot propagate indefinitely along the waveguide at any arbitrary angle, even if it meets this condition.  Rather, only those angles whose wavefronts add together in phase will be supported by the waveguide.  These define the modes of the waveguide, and the idea can be expanded to off-axis helical rays, and other types of non-step-index waveguides.  Another (equivalent) way to understand modes is to consider the electric (or magnetic) field distribution across the face of the waveguide.
Within the waveguide core there will exist a number of supported waves, which vary approximately sinusoidally in radius and azimuth; in the cladding the electric field will exponentially decay.  The exact form of the electric field distribution will depend on the waveguide geometry, and the number of supported modes will depend on the geometry, the NA, the waveguide width, and the wavelength of light.
For example, for a circular step index fibre, separating the electric field into radial ($r$), azimuthal ($\phi$), and axial ($z$) components gives,
\begin{equation}
   E(r,\phi,z)  = E_{lm}(r) \cos l\phi\ {\rm e}^{{\rm i}(\omega t - \beta z)},
\end{equation}
\begin{eqnarray}
E_{lm}(r) & \propto &
\begin{cases}
    J_{l} (k_{\rm T_{m}}) & r \le a \\
    k_{l} (\gamma_{m} r)  & r > a
\end{cases} \\
k_{\rm T_{m}} &= & n_{\rm core}^{2} k_{0}^{2} - \beta_{m}^{2},\\
\gamma_{m}^{2} &= &\beta_{m}^{2} - n_{clad}^{2} k_{0}^{2}
\end{eqnarray}
where $a$ is the fibre radius and $k_{0}$ is the wavenumber in vacuum.  Figure~\ref{f:fibremodes} shows the electric field distribution for the first 9 modes of a fibre with $a=50$~$\mu$m, $n_{\rm core}=1.51$ and $n_{\rm clad}=1.52$ at $\lambda_{0} = 1.55$~$\mu$m.

\begin{figure}
    \centering
    \includegraphics[scale=0.8]{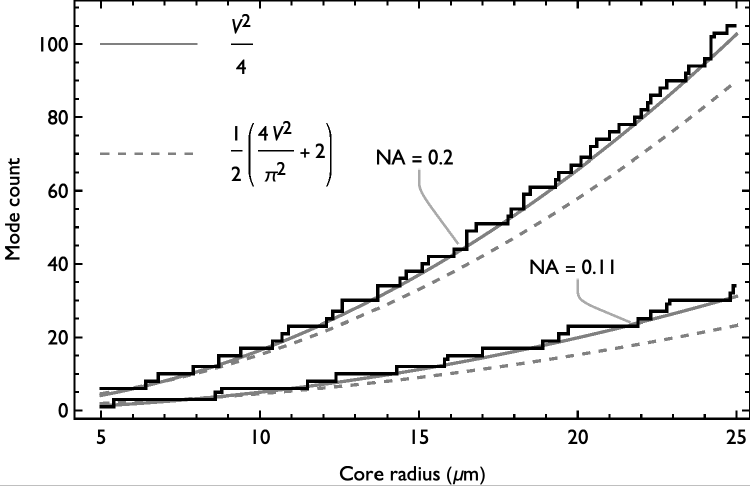}
    \caption{The number of modes in a step-index fibre (thick black lines) for fibres with NA = 0.11 and NA = 0.2 as a function of core-radius, compared to approximations.  Note tha the $V^{2}/4$ approximation is more accurate at high mode count.}
    \label{f:modecount}
\end{figure}

The number of modes in a fibre, neglecting polarisation, may be approximated as 
\begin{equation}
\label{e:nmodes}
  M \approx V^{2}/4  ,
\end{equation}
where $V$ is the normalised frequency,
\begin{equation}
\label{e:vpar}
V = 2 \pi \frac{a}{\lambda_{0}}\ N\!A.
\end{equation}
Note, this approximation is more accurate than the commonly used $M \approx \frac{1}{2}\left(\frac{4 V^{2}}{\pi^{2}} + 2 \right)$, see Figure~\ref{f:modecount}. A step-index fibre can support only one mode if $V<2.405$.

Other types of waveguides operate on the same principles.  Lithographic waveguides have a rectangular cross-section core made from a high-index material such as indium phosphide, silicon nitride, or silicon, surrounded by a lower index material cladding, such a silica, PMMA (polymethyl methacrylate), or air.  Likewise, direct-write waveguides, in which a waveguide is written into a block of glass with a high-powered laser thereby modifying the refractive index of the glass to create a high-index core. Photonic crystal fibres create a low index cladding from the same material as the fibre core but punctuated with an array of air-holes running along the length of the fibre.  Graded-index fibre operates on similar principles, but the change in refractive index is smooth, causing the light to follow a smooth curved path along the waveguide core, and allowing for better dispersion control.

Some types of waveguides do not operate on the principle of total internal reflection.  Photonic bandgap fibre has a cladding consisting of an array of holes, similar to photonic crystal fibre, but in this case the structure of the holes prevents light of certain wavelengths propagating into the cladding, and thus confining the light to the core.

\subsection{Coupling into waveguides}

A crucial consideration for astrophotonics is the coupling of light from a telescope into a waveguide.  In the extreme case this can mean coupling the light from a 8~m primary mirror, or greater, into a fibre of diameter $<10$~$\mu$m, or in the case of integrated photonics into waveguides $<1$~$\mu$m wide.  This is not a trivial task.  

\begin{figure}
    \centering
    \includegraphics[scale=0.4]{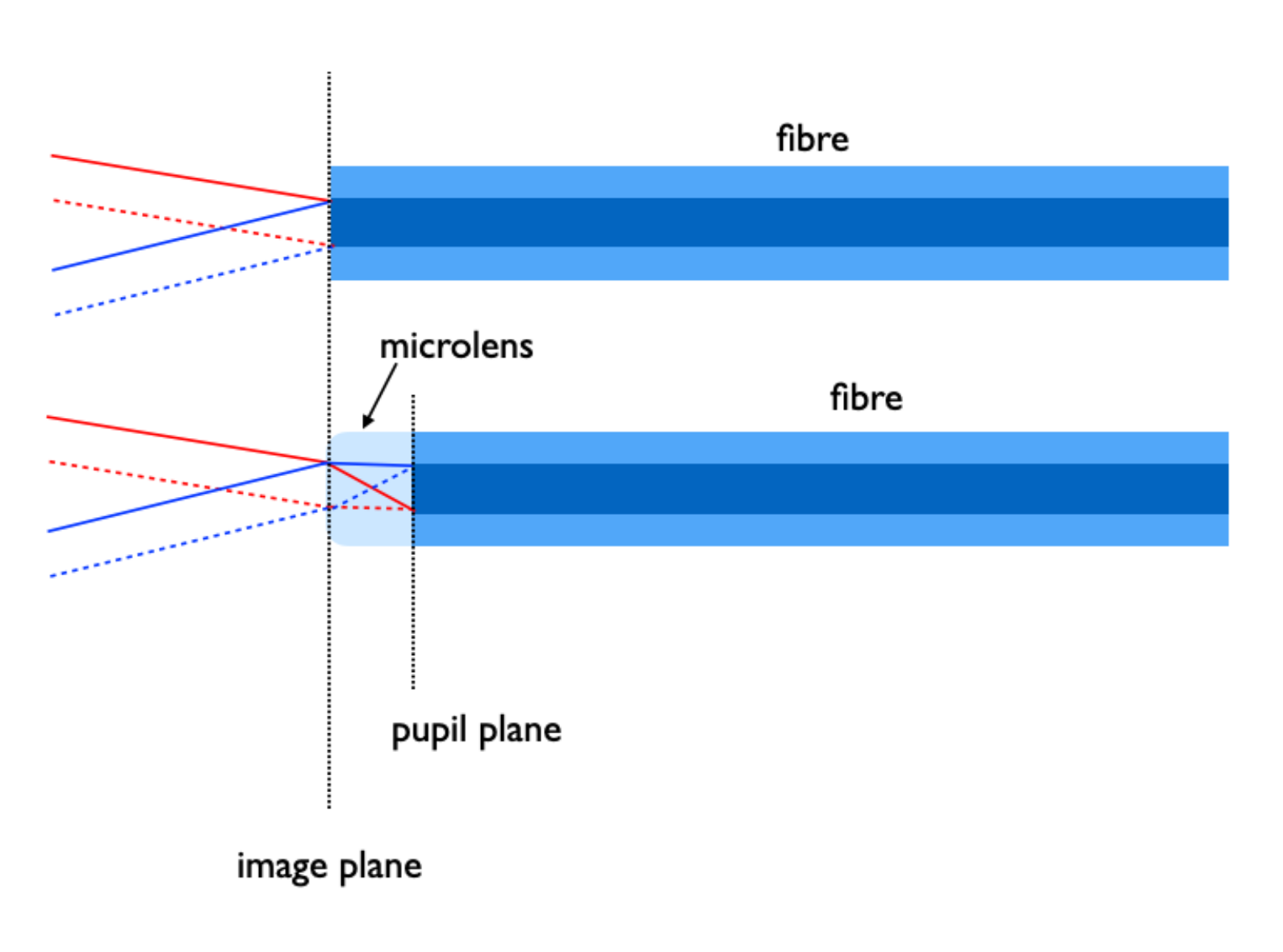}
    \caption{Sketches of the fibre coupling schemes, using bare fibre to couple an image into the fibre (top), or using a microlens to couple a pupil image into the fibre (bottom).}
    \label{f:fibrecouple}
\end{figure}

There are two common schemes for coupling light into a waveguide, see Figure~\ref{f:fibrecouple}.  First, the end face of the waveguide can be put at an image-plane of telescope or instrument, and thus a focussed image of (part of) the astronomical object is formed on the face of the waveguide.  Secondly, a microlens can be attached to the end of the fibre; the image is formed on the face of the microlens, which then projects an image of the telescope pupil onto the face of the fibre.  This latter scheme has some advantages over the former, at the cost of increased complexity, (i) the input speed of the beam can be matched to the fibre NA, or else the spot size can be matched to the fibre core diameter, (ii) microlenses can be tessellated into an array to allow a contiguous field-of-view for spatially resolved spectroscopy, (iii) the pupil image will approximate a top-hat profile, providing a  good match to the approximately top-hat mode profile of a multimode fibre.

The efficiency with which light can be coupled into a waveguide is given by the overlap integral of the input electric field distibution and the electric field distribution of the modes of the waveguide, i.e.,
\begin{equation}
\label{e:modeolap}
    \eta = \sum_{m=1}^{M} \frac{\left| \bigintss\! \! \! \bigintss E_{0}E^{*}_{m}\ {\rm d}x\ {\rm d}y\right|^{2}}{\bigintss\! \! \! \bigintss \left|E_{0}\right|^{2} {\rm d}x\ {\rm d}y\ \bigintss\! \! \! \bigintss \left|E_{m}\right|^{2} {\rm d}x\ {\rm d}y}, 
\end{equation}
where $E_{0}$ is the input electric field, $E^{*}_{m}$ is the complex conjugate of the electric field of the $m$th mode, and there are $M$ modes in total.  The denominator ensures that the integral is normalised such that the efficiency $0\le \eta \le 1$.

The mode overlap given by equation~\ref{e:modeolap} becomes very important when dealing with single or few-mode waveguides.  When considering waveguides carrying a large number of modes we can ensure high  coupling efficiency by conservation of etendue., i.e., setting
\begin{equation}
\left(a\ N\!A\right)^{2} \approx \left(\frac{D_{\rm tel} \Gamma}{4}\right)^{2}
\end{equation}
where $\Gamma$ is 
angular size of the object being observed in radians,
and $D_{tel}$ is the diameter of the telescope.  This leads to a condition on the number of modes required via equations~\ref{e:nmodes} and \ref{e:vpar}, such that the fibre must support at least
\begin{equation}
\label{e:modesee}
M_{\rm see} \approx \left( \frac{\pi D_{\rm tel} \Gamma}{4 \lambda}\right)^{2}.   
\end{equation}

In the single and few-mode limit we must compute the mode overlap from equation~\ref{e:modeolap}.  In general we may describe the input electric field in the pupil plane by
\begin{equation}
    E_{\rm pupil}(r) = E_{\rm S}(r) P(r) \Psi(r),
\end{equation}
where $E_{\rm S}$ is the electric field amplitude of the source, $P(r)$ is the pupil transmission function, and $\Psi(r)$ is a turbulent phase screen.  In the image plane the electric field is given by the Fourier transform of this, i.e.,
\begin{equation}
    E_{\rm image}(r) = \hat{E}_{\rm pupil}\left(\frac{r}{\lambda F D_{\rm tel}}\right),
\end{equation}
where $F$ is the focal ratio at the image plane.  

If the angular resolution is limited only by the diffraction of the telescope aperture, i.e. the so-called diffraction limit, i.e. $\Psi(r) = 1$, we have,
\begin{equation}
     E_{\rm image}(r) = E_{\rm s}\left(\frac{2 J_{1}(s)}{s} - \alpha^{2} \frac{2 J_{1}(\alpha s)}{\alpha s}\right),
\end{equation}
where $\alpha D_{\rm tel}$ is the diameter of the telescope central obstruction.  For a single-mode fibre (SMF) in the diffraction limit the maximum coupling efficiency is $\approx 80$\% \cite{sha88}.  Horton \& Bland-Hawthorn\cite{hor07} have calculated the coupling efficiencies for few mode fibres in the diffraction limited regime.  These calculations have been extended to the pupil plane and seeing limited conditions, i.e. when the angular resolution is degraded by the distortion of the incoming wavefronts by atmospheric turbulence.\cite{ell21,cha21,ell23}.

In anything other than nearly diffraction limited conditions the coupling efficiency into single mode fibre is very low, and for seeing limited conditions the use of highly multimode fibres is necessary; see equation~\ref{e:modesee}.  However, many photonic functions rely on single mode waveguides for correct operation (since the phase must be precisely controlled).  Therefore, in seeing limited conditions it is necessary to couple into multimode fibre and thereafter to couple to multiple replicated single mode devices.  

The photonic lantern\cite{leo05,bir15} was invented to solve this specific problem.  Photonic lanterns work on the principle of tapering fibres, such that a larger fraction of the electromagnetic field is in the evanescent field, and less is confined to the core.  If a bundle of $N$ single mode fibres is surrounded by a low index jacket, and made to undergo a slow taper, eventually all the light will be contained in the evanescent field.  The $N$ supermodes will be guided by the low index jacket, effectively forming a multimode fibre.  The process is necessarily reversible, which allows a multimode fibre to convert to $N$ single modes and vice versa.  The photonic lantern is a cornerstone of astrophotonics, allowing single mode photonic devices to be employed on ground-based telescopes working in turbulent seeing-limited conditions.

Alternatively one can first correct the turbulent wavefronts using adaptive optics (AO) to allow efficient coupling directly into single mode waveguides.   The best coupling efficiency to date is $>40$\% \cite{jov17}, cf. to a theoretical coupling efficiency of $\sim 80$\% \cite{sha88}. This technique will be even more important in the future era of extremely large telescopes (ELTs) with very high AO correction.  However, current AO techniques suffer from limited sky-coverage due to the necessity for natural guide stars to correct for non-common path aberrations in laser guide star systems.

A comparable difficulty arises in coupling from single mode fibre into very narrow high index channel waveguides, such as Si or Si$_{3}$N$_{4}$.  These waveguides have cross-sections of 0.25 -- 1~$\mu$m, and therefore the mismatch in mode-field diameter with a SMF of $\sim 10$~$\mu$m diameter leads to very large losses ($\approx 13$~dB).  One solution is to use an inverted taper\cite{mcn03,alm03,fu14,tie15} in the channel waveguide to expand the mode-field diameter to match the SMF.  This has the advantage of being polychromatic, but tolerances on the alignment and mode matching are very tight and difficult to achieve.  Another solution is to use grating couplers\cite{tai03,tai04},  which can achieve efficiencies of $> -1$~dB ($> 80$ per cent).  However, these efficiencies peak in a narrow range of wavelengths.  Typical 3~dB bandwidths are $\approx 40$ -- 70~nm, (cf.\  typical astronomical passbands of $\sim 300$~nm).

\subsection{Combining and dividing light}
\label{sec:combine}

One of the most important and successful uses of astrophotonics, beyond the (photonically) simple application of transporting light with fibres, is to combine light from different waveguides, or equivalently to divide light from a single waveguide into two outputs.  These techniques are especially beneficial for interferometric applications (see \S~\ref{sec:interferometer}).  

\begin{figure}
\centering
\includegraphics[scale=0.3]{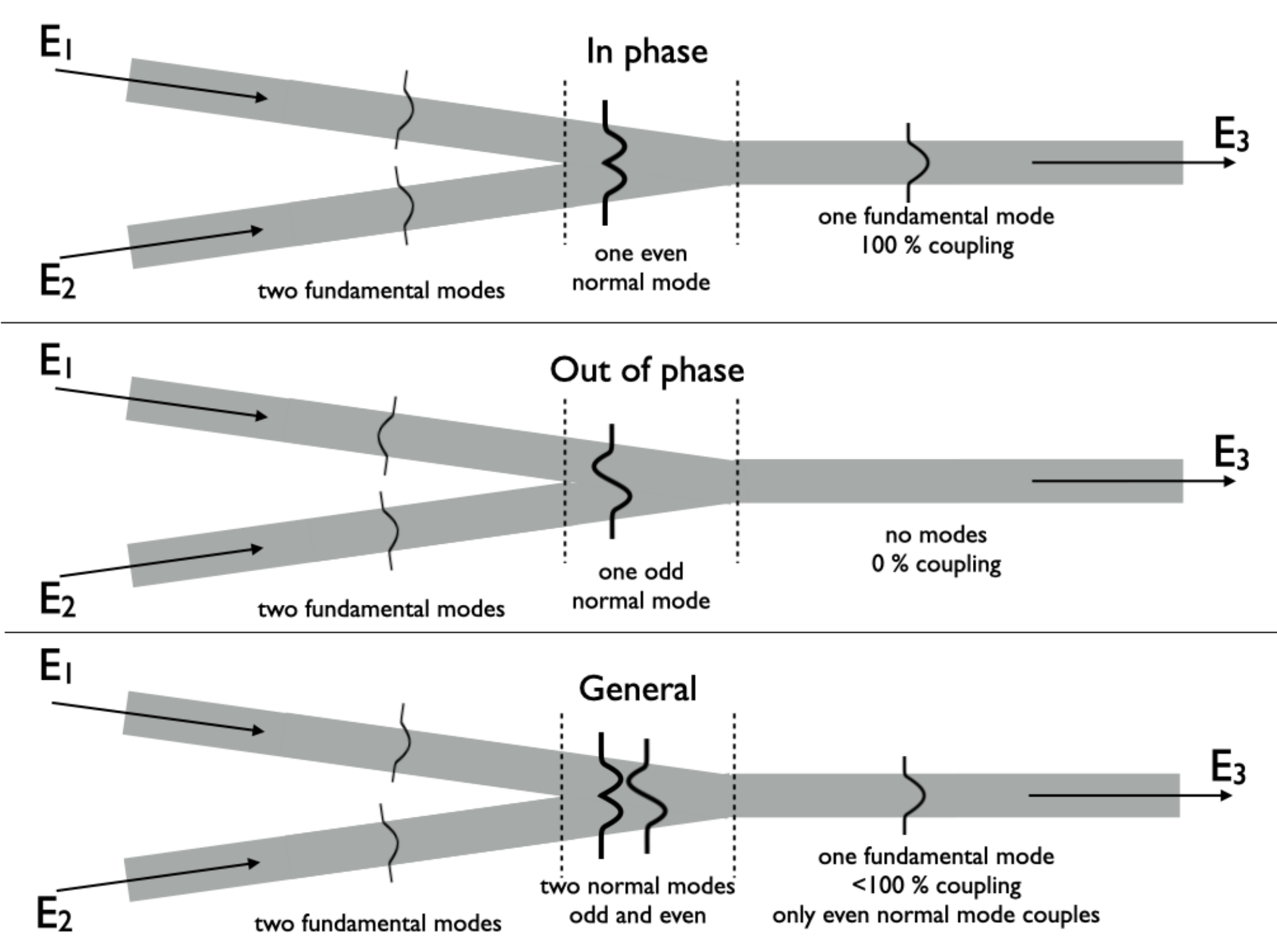}
\caption{Schematic diagram of a photonic Y-branch beam combiner, consisting of two single-mode waveguides combining into one.  The power in the outgoing mode depends on the relative phase of the incoming modes.}
\label{f:pbc}
\end{figure}

Light can be combined from two waveguides into one using a Y-branch coupler as sketched in Figure~\ref{f:pbc}.  However the two input waveguides will not necessarily couple all their power to the output waveguide: in general the coupling will be lossy.  It is only possible to couple all the power into a single outgoing waveguide when the two inputs are exactly coherent (top panel of Fig.~\ref{f:pbc}).  If the light in the two input waveguides is exactly out of phase then no light will couple to the outgoing waveguide; it will all be radiated and lost (middle panel of Fig.~\ref{f:pbc}).

The general situation can be understood by considering that the two incoming modes will excite exactly two supermodes.  However, the outgoing waveguide can only carry one mode, viz.\ the even normal mode; the higher order odd normal mode will be lost.  The final power will therefore depend on the relative phase of the incoming modes, or equivalently on the relative power in the even and odd modes.
Interestingly, if light is input into only one of the incoming waveguides, then both even and odd normal modes are equally excited, and the coupled power can not exceed 3~dB for a symmetric coupler.

The reverse situation of a Y-branch splitter will split the light from an incoming waveguide equally into two outputs, for a symmetric device.  Asymmetric splitting can be arranged with asymmetries in the output waveguides, to control the fractional power or polarisation splitting etc.

\begin{figure}
\centering \includegraphics[scale=0.8]{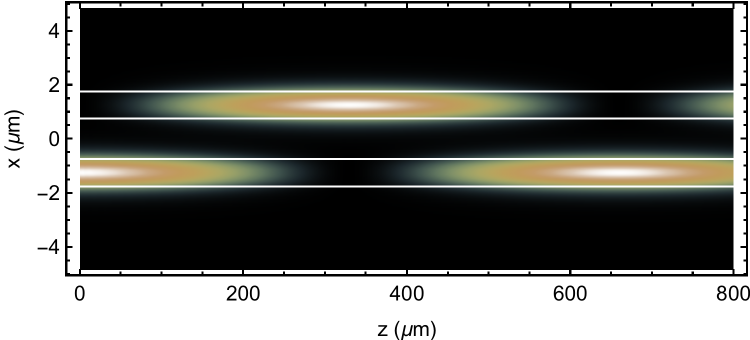}
\caption{The power (indicated by the colour) coupled between two parallel Si$_{3}$N$_{4}$ waveguides with SiO$_{2}$ cladding as a function of distance along the waveguides for light at a wavelength of 1550~nm.  }
\label{f:directionalcoupler}
\end{figure}

It is possible to couple from one waveguide to an adjacent waveguide when the two waveguides are close enough, such that the evanescent field from the first can overlap with the core of the second, see Figure~\ref{f:directionalcoupler}.

The power coupled between the waveguides can be derived using supermode analysis\cite{ell23}. 
The power in each core, $P_{1}$ and $P_{2}$, oscillates sinusoidally,  
 \begin{eqnarray}
P_{1} &=& \cos^{2} C z, \\
P_{2} &=&  \sin^{2} C z,\label{e:powcoupwgs}
\end{eqnarray}
where $z$ is the length along the optical axis, and $C$ is the coupling coefficient,
\begin{equation}
C_{jm} = \frac{k}{2 n_{0}} \int_{A} \widehat{\psi}_{j} \widehat{\psi}_{m} \Delta n_{m}^{2} dA\label{e:7mcf_coup}
\end{equation}
where $\widehat{\psi}_{j}$ and $\widehat{\psi}_{m}$ are the modes of each waveguide, $k$ is the wavenumber, 
\begin{equation}
\Delta n_{j}^{2}(x,y) = 
\begin{cases}
n_{\rm co}^{2} - n_{0}^{2}, & {\rm in\ each\ core}\ j \\
0, & {\rm elsewhere.}
\end{cases}
\end{equation}
and $n_{0}$ is the refractive index of the surrounding material. 
The power oscillates between the two cores, with a beat length 
\begin{equation}
L_{\rm C} = \frac{\pi}{C}
\end{equation}
Note that by controlling the length of the coupling region, any fraction of power can be coupled into the second waveguide.

\begin{figure}
    \centering
    \includegraphics[scale=0.35]{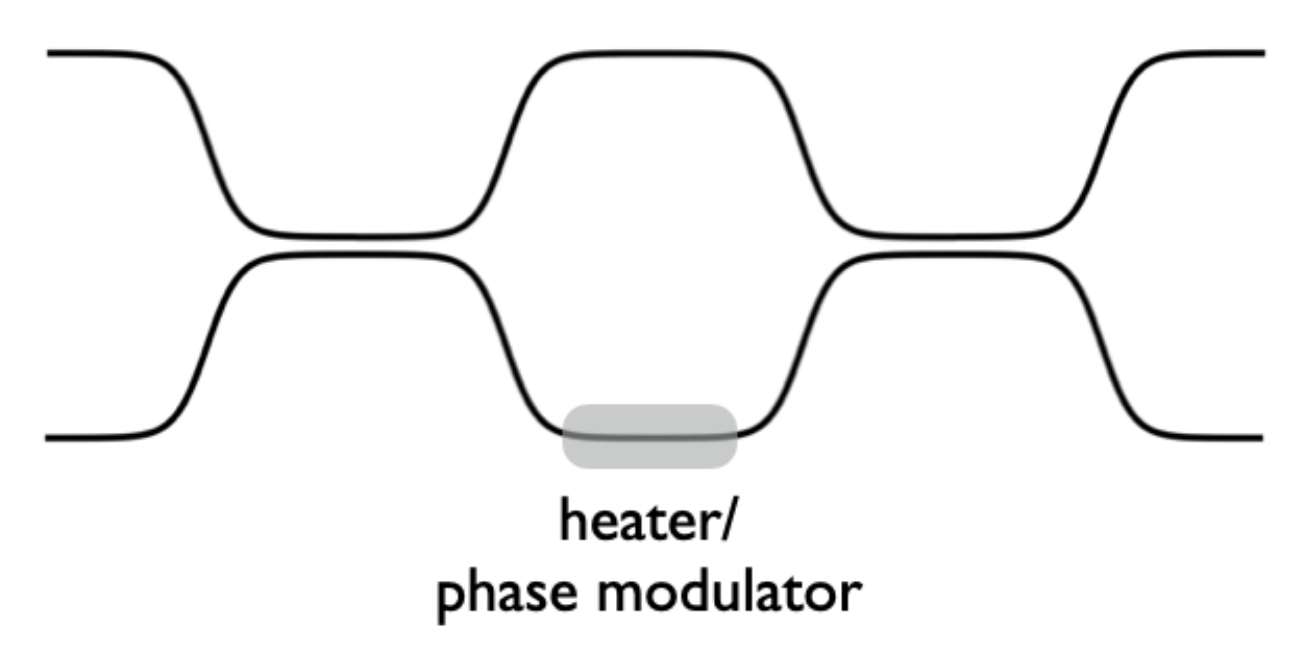}
    \caption{Sketch of an integrated photonic Mach-Zehnder coupler.}
    \label{f:mz}
\end{figure}

More control over the coupling fraction can be obtained by extending this method to a Mach-Zehnder coupler as sketched in Figure~\ref{f:mz}.   In this case, light from a single input is split equally between two waveguides with a 50/50 evanescent coupler, whereupon one of outgoing waveguides has a phase shift applied to it, typically from an embedded heater.   The two waveguides are then coupled again, but now the output into each waveguide can be modulated by controlling the phase shift, allowing either the precise combination of light from the two input waveguides, or the precise division of light from a single input.

\subsection{Dispersing light}
\label{sec:disperse}

Spectroscopy is a fundamental part of astrophysics, and so the ability to disperse light is critical.  Classical astronomical instruments use prisms or diffraction gratings of various types for this purpose.  Analogous methods exist in photonic circuits, which we describe below.  Similarly, many photonic devices use wavelength division multiplexing (WDM) to combine and separate different signals encoded at different wavelengths.  In the latter case, however, it is usually only necessary to separate and measure discrete signals, whereas in the former a continuous spectrum is  required.  We will discuss WDM further in the next section on filtering (\S~\ref{sec:filter}), and here we describe photonic methods to obtain continuous spectra.

An important property of all the approaches described in this section is that they fundamentally break the scaling of instrument size with telescope size, which applies to conventional optical spectrographs.  Instead, the problem of larger optics and instruments is replaced by one of replication; many identical single mode photonic spectrographs replace one large monolithic spectrograph.  This property becomes more apposite as we enter the era of extremely large telescopes.  This will be discussed further in section~\ref{sec:microspect}.

The simplest scheme is to follow the same scheme as for ordinary fibre-fed spectrographs, but to feed the optics with single-mode fibres.
Although seemingly trivial, this change can lead to important consequences.  
First, the optics of the spectrograph can be minimised to the smallest size for a fixed resolution.  Secondly, the point spread function of the spectrograph shows remarkably little scattering since the high spatial frequencies are suppressed due to the Gaussian spatial mode profile of the SMF\cite{bet13}.

\begin{figure}[t]
    \centering
    \includegraphics[scale = 1.4]{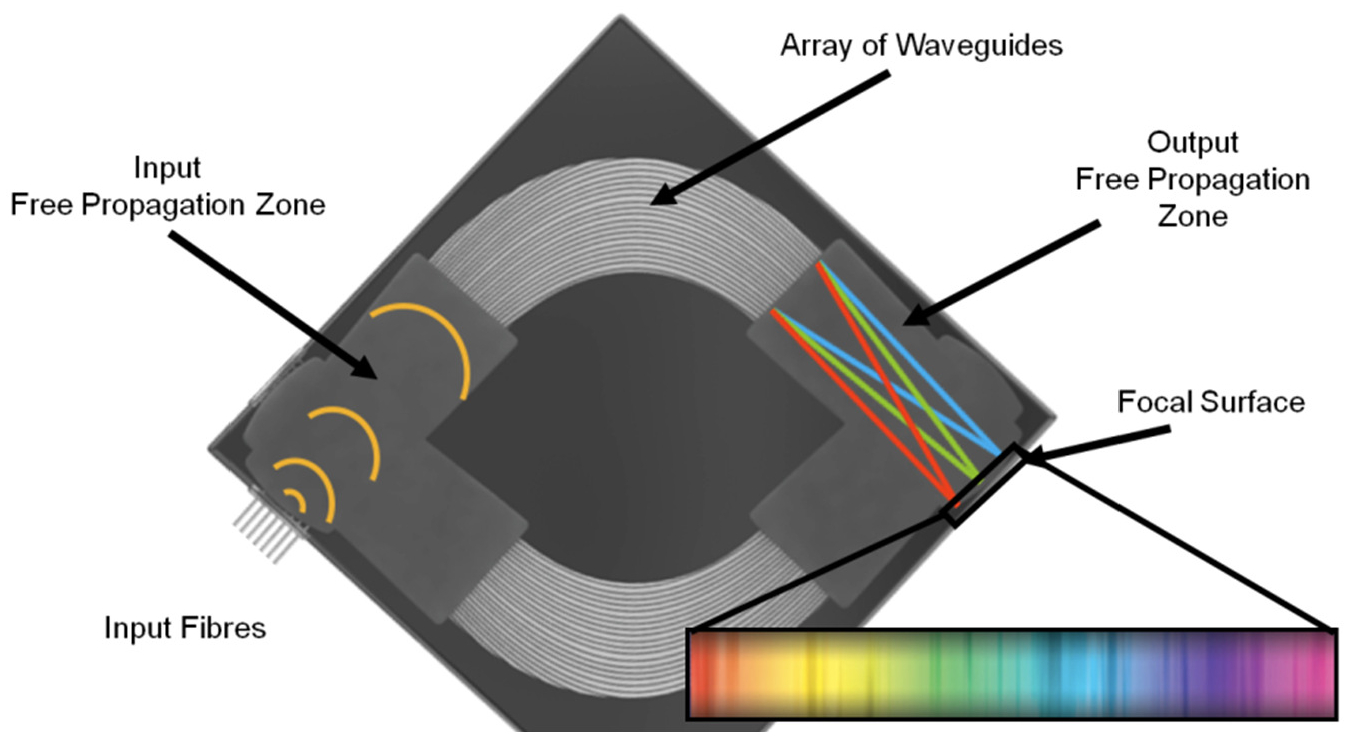}
    \caption{Schematic diagram of an arrayed waveguide grating.  The device actually contains two AWGs only the top one of which has been shown connected to input fibres\cite{cve12b}.  
    }
    \label{f:awg}
\end{figure}

These ideas can be extended to replace the spectrograph optics themselves with photonic circuits.  The most studied of these are arrayed waveguide gratings (AWGs), the components of which are shown in Figure~\ref{f:awg}.  Light enters through one or more input SMFs, whereafter it propagates unguided through a free-propagation zone.  It is then coupled to an array of waveguides, each of which has a fixed offset in path length from its neighbour.  This change in path length is responsible for the change in phase which results in interference at the output of the waveguide array.  The light can be refocussed onto an output array of fibres, or on to a surface by propagating unguided across another free propagation zone symmetrical to the first.
The use and development of AWGs for astronomy will be discussed further in section~\ref{sec:microspect}.

\begin{figure}
    \centering
    \includegraphics[scale=0.3]{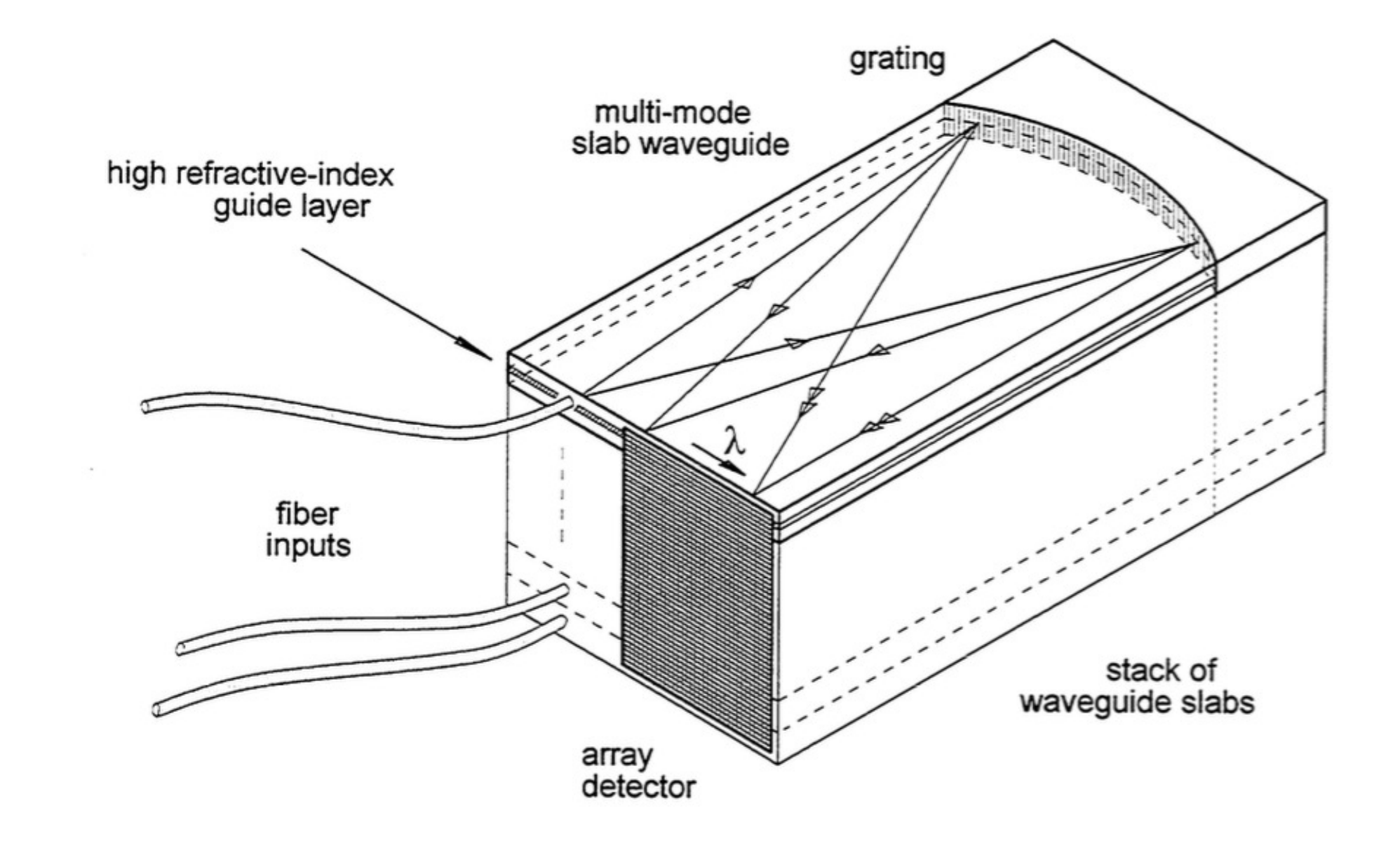}
    \caption{Simplified sketch of a photonic echelle spectrograph illustrating the concept\cite{wat95}.  Light delivered from an input fibre propogates freely across a slab waveguide to the photonic echelle grating surface.  The reflected light interferes and forms a spectrum as for a conventional echelle.   The output is usually coupled into discrete fibres, but for astronomical purposes these can be dispensed with, as shown here, to obtain a continuous spectrum.  A stack of such devices could be made to form a miniature multi-object spectrograph.}
    \label{f:peg}
\end{figure}

Another photonic on-chip solution is the photonic echelle grating (PEG).  These have received limited attention for astronomical applications\cite{wat95,wat96,xie18,sto19} since they are less well developed for other WDM applications, although they may in fact have some advantages, especially in terms of smaller footprint, less cross-talk (and hence fewer losses), and improved finesse.  The operation of a PEG can be seen in the simplified conceptual sketch from Watson (1995)\cite{wat95} in Figure~\ref{f:peg}.  An input waveguide  deliveres light to a free propagation region, after which it is reflected by the echelle grating.  This is similar to a conventional echelle, but is made by etching the surface into the photonic layer through photo- or electron-beam lithography, after which the facets can be coated, e.g. with silver.

A rather different scheme for a micro-spectrograph is to use a Lippmann spectrograph, or stationary wave integrated Fourier transform spectrogpah (SWIFTS)\index{SWIFTS}\index{Stantionay wave integrated Fourier transform spectrograph}, embedded into a waveguide.  In this technique light is either injected into a single-mode waveguide and reflected at one end, or is injected into both ends of a single-mode waveguide.  In either case, this results in two identical and counter-propagating waves, which will construct a standing wave inside the waveguide.  This standing wave can be detected by placing nano-dots along the outside of the waveguide which scatter the evansecent field onto a detector, recording an interferogram.  The interferogram can then be analysed in a similar way to a Fourier transform spectrometer in order to recover the spectrum\cite{con95,lec07,bli17}.

To make any of these photonic spectrographs with acceptable efficiency requires more than a single SMF input. The coupling efficiency into a SMF is poor in all but nearly diffraction limited conditions, which is only possible from the ground with extreme adaptive optics\cite{sha88,jov17,ell21}.  Secondly, in order to capture sufficient \'{e}tendue in seeing limited conditions a multimode fibre is required (equation~\ref{e:modesee}).  Therefore, a photonic lantern can be used to capture the light, before splitting into parallel single mode fibres, which can form the pseudo-entrance-slit to the diffraction limited spectrograph.

\subsection{Filtering}
\label{sec:filter}

Filtering, either to isolate a specific signal, or to remove a source of noise, is an important part of astronomy.  Photonic  filters (i.e.\ filters embedded into waveguides) offer much more complex filtering than traditional absorptive, interference, or holographic filters, with the ability to isolate multiple ($>100$) specific signals in a single waveguide.   Many photonic filters have been developed for wavelength division multiplexing applications, such that multiple signals can be transmitted down the same waveguide, and accurately separated upon receipt.  The challenge for astrophotonics is to achieve the same thing for hundreds of signals, spread over a very large bandwidth ($\approx 300$~nm) and in multimode fibre.

To date, the only astrophotonic filters  to have been incorporated in an astronomical instrument are fibre Bragg gratings\cite{bland04,bland08,bland11}, but there are many alternatives under development (see chapter 10 of \cite{jov23}).  Many of these are based on the principle of Bragg gratings, whether written into fibre, glass or photonic circuits, and other are based on resonators.

\begin{figure}
    \centering
    \includegraphics[scale=0.3]{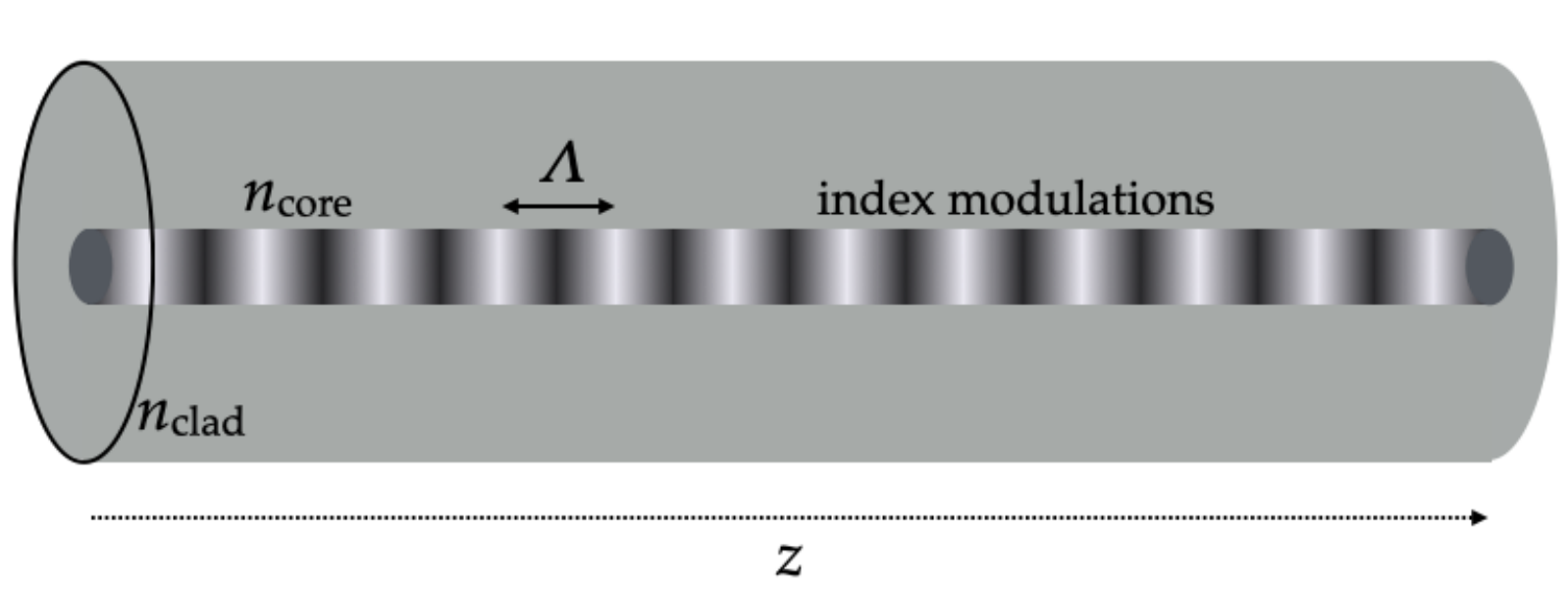}
    \caption{Sketch of a simple fibre Bragg grating showing the variation in the refractive index of the core with period $\Lambda$.}
    \label{f:fbgsketch}
\end{figure}

Bragg gratings operate on the principle of Fresnel reflection.  Consider a periodic variation in the refractive index of the core of a fibre, as sketched in Figure~\ref{f:fbgsketch}.  As light propagates down the fibre a small fraction is reflected backwards at each change in refractive index.  For light at wavelengths meeting the condition $m \lambda_{\rm B} = 2 n_{\rm core} \Lambda$, where $m$ is an integer, $\Lambda$ is the pitch, and $n_{\rm core}$ is the average refractive index of the core, the reflected light will constructively interfere, leading to a strong reflection at wavelengths $\lambda_{\rm B}$.  More complex filters can be made by using aperiodic modulations in the index; in this way fibre Bragg gratings (FBGs) have been made with 150 notches precisely matched in wavelength, strength and width to the target signals\cite{bland08}.

The refractive index variations can be inscribed into the waveguides in various ways.  The most common is to use a UV laser interferometer to irradiate the core (often doped and hydrogenated) to alter the refractive index.  This can also be done in a point-by-point method using the direct-write technique\cite{mar06,zha07} with a femto-second laser, either into a fibre\cite{goe18} or into a block of glass\cite{spa14}.  For waveguides on a photonic chip the effective index can be modulated by changing the width of the waveguide.

\begin{figure}[t]
\begin{center}
\subfigure[]{
\includegraphics[scale=0.25]{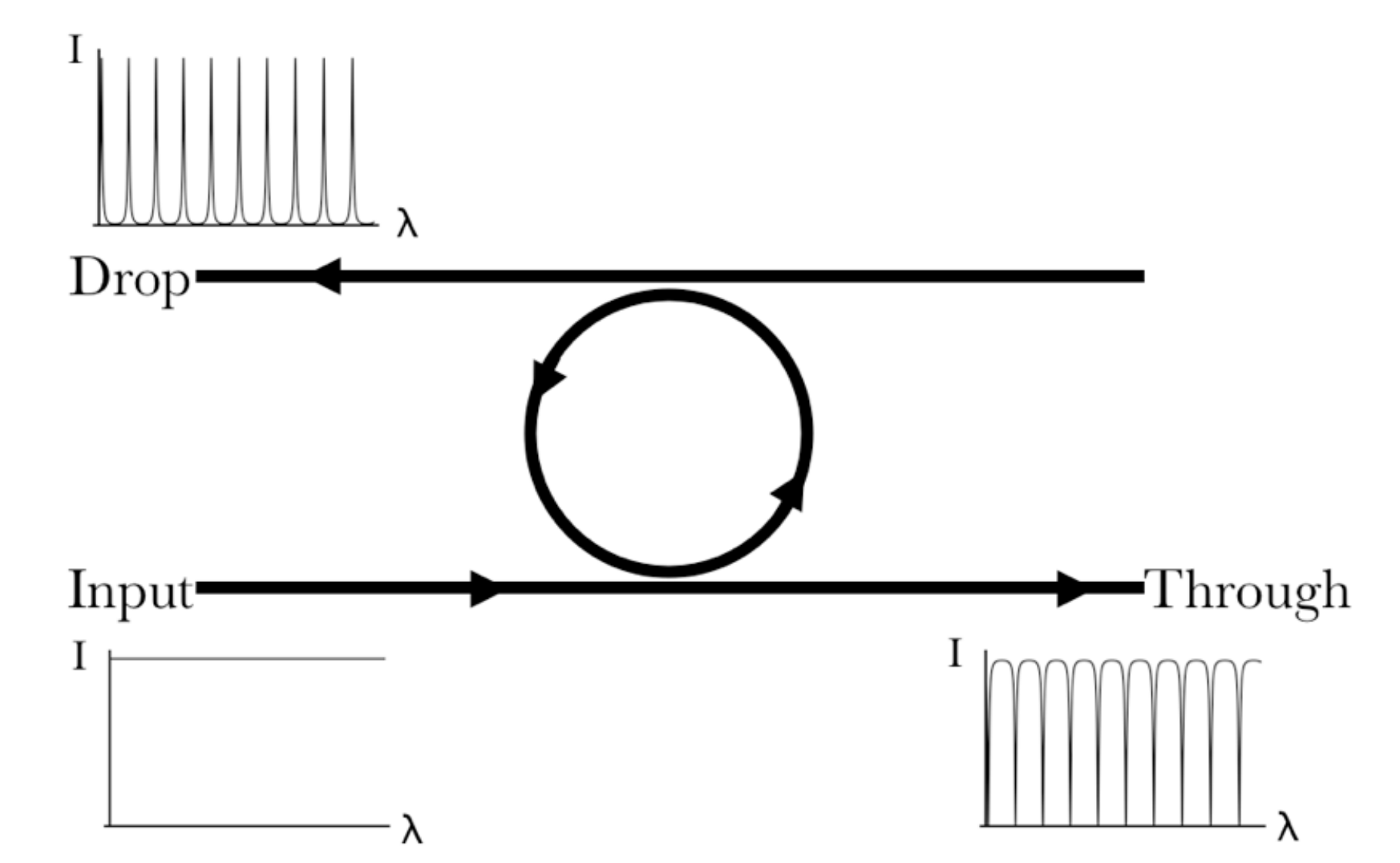}
}
\subfigure[]{
\includegraphics[scale=0.3]{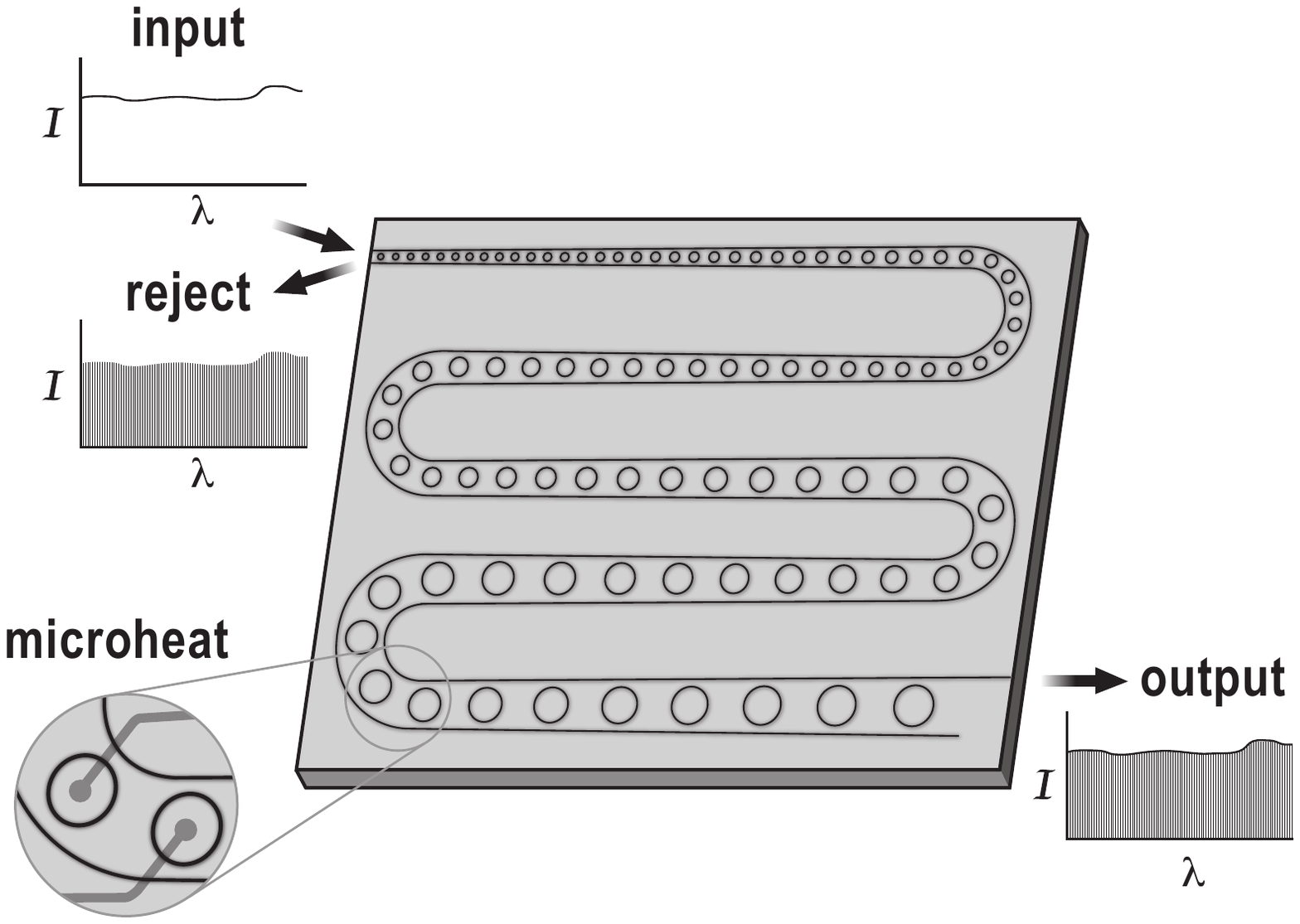}
}
\caption{Schematic depictions of ring resonators for use as either notch filters, narrow bandpass filters, or to generate a comb of frequencies.}
\label{f:ringres}
\end{center}
\end{figure}

An alternative to using Bragg gratings is to use a resonator to filter specific signals.  For example, consider Figure~\ref{f:ringres}a.  Light from the input waveguide evanescently couples to the ring waveguide.  If the optical path length around the ring is equal to an integer number of wavelengths, i.e. $m \lambda = n L$, where $n$ is the effective index and $L$ is the length, then light at that wavelength will constructively interfere.  This light couples back into the input waveguide exactly out of phase with the incoming light, leading to a series of notch filters at the resonant frequencies of the ring.  Light can also be coupled into a separate drop port leading to a series of peaks at the resonant frequencies.  A series of rings, as in Figure~\ref{f:ringres}b can be used to selectively filter multiple features\cite{ell12c,ell17,liu21}.




\subsection{Quantum astrophotonics}
\label{sec:qm}

One of the practical differences between astrophotonics and classical optics is in the description of light.  For some classical instruments a geometrical optics description may be adequate; if it is necessary to also include phase information then an electromagnetic description may be necessary taking into account fields, polarisation etc.  The ultimate description of light is quantum mechanical, in which the state of individual photons are taken into account.  

A quantum mechanical understanding of light allows new ways of building instruments that can exploit quantum mechanical phenomena.  We describe two such proposals for very long baseline optical interferometry in section~\ref{sec:quantuminter}.   Here we introduce some of the prerequisite quantum photonic functions and background.

\subsubsection{Quantum states and qubits}

The state of a photon may be described in terms of its orbital angular momentum and 3 components of the linear momentum vector, or equivalent quantities such as its polarisation.  
%

A two-state system, for example horizontal or vertical polarisation of a photon, can describe a qubit, the quantum mechanical equivalent of a bit.  Unlike a classical bit, which must be in one state or the other, a qubit can be in a coherent superposition of the two states simultaneously.  A second important distinction is that measurement of a classical bit does not disturb its state, whereas measurement of a qubit would always disturb the superposition.

The general quantum state of a qubit can be represented by a linear superposition of 
two 
basis states (e.g.\ the polarisation states).  Let these be denoted as,
\begin{equation}
\label{eqn:0q}
\Ket{0} = 
\begin{bmatrix}
1 \\
0
\end{bmatrix},
\ \ \ 
\Ket{1} = 
\begin{bmatrix}
0 \\
1
\end{bmatrix}.
\end{equation}

This means that a single qubit, $\psi$, can be described by a linear combination of $\ket{0}$ and $\ket{1}$, i.e.,
\begin{eqnarray}
\Ket{\psi} = \alpha\Ket{0} +\beta \Ket{1}, \\
    |\alpha|^{2} + |\beta|^{2} = 1.
\end{eqnarray}
where $\alpha$ and $\beta$ are probability amplitudes.


\subsubsection{Entanglement}

Two particles, or two qubits, can be entangled when they are produced in the same event.  Entanglement means that the measurement of the state of one qubit entirely determines the state of the other qubit.

\paragraph{Formal definition}

A two-qubit state can be made via a product of two single qubit states (technically, this is a Kronecker tensor product), which is given by
\begin{equation}
    \begin{pmatrix} a_1 \\ a_2 \end{pmatrix} \otimes \begin{pmatrix} b_1 \\ b_2 \end{pmatrix} = 
    \begin{pmatrix} a_1 b_1 \\ a_1 b_2 \\ a_2 b_1 \\ a_2 b_2 \end{pmatrix}
\end{equation}
such that

\begin{equation}
    \vert 00\rangle = 
    \begin{pmatrix} 1 \\ 0 \\ 0 \\ 0 \end{pmatrix}, \ \ \ \
     \vert 01\rangle =
    \begin{pmatrix} 0 \\ 1 \\ 0 \\ 0 \end{pmatrix}, \ \ \ \ 
     \vert 10\rangle =
    \begin{pmatrix} 0 \\ 0 \\ 1 \\ 0 \end{pmatrix}, \ \ \ \
     \vert 11\rangle = 
    \begin{pmatrix} 0 \\ 0 \\ 0 \\ 1 \end{pmatrix}.
\end{equation}

\paragraph{How to entangle single qubits}
\label{sec:entangle}

Photons can be entangled into two-qubit states, as will be described below.  Here we describe the mathematics of this entanglement, using operators to manipulate the states.  The most important operator
is the $\boxed{\mathbf{CNOT}}$ gate. The gate receives two qubits, a control and a target qubit; it also returns two qubits but in an entangled state. The operator rules are that if the control qubit is $\vert 0\rangle$, the target qubit is untouched; otherwise $\vert 0\rangle$ and $\vert 1\rangle$ are flipped, i.e. the operator performs an X-gate operation. The operator matrix follows:
\begin{equation}
\boxed{\mathbf{CNOT}} =
    \left( \begin{array}{cccc}
    1 & 0 & 0 & 0 \\
    0 & 1 & 0 & 0 \\
    0 & 0 & 0 & 1 \\
    0 & 0 & 1 & 0 
    \end{array} \right) 
    \label{eqn:cnot}
\end{equation}
The CNOT gate operates on all four basis states of a two-qubit system as a superposition, i.e. simultaneously. 
The gate is fully reversible in order to determine the original qubit states.

Bell states are four maximally-entangled quantum states of two qubits, which are given by,
\begin{eqnarray}
\vert \psi_+ \rangle = {\rm CNOT}\; \mathbf{H}\otimes\mathbf{I}\; \vert 00\rangle& = &
    \frac{1}{\sqrt{2}} \left( \begin{array}{cccc}
    1 \\ 0 \\ 0 \\ 1 
    \end{array} \right) \\
  \vert \phi_+ \rangle =  {\rm CNOT}\; \mathbf{H}\otimes \mathbf{I}\; \vert 01\rangle &=& 
    \frac{1}{\sqrt{2}} \left( \begin{array}{cccc} 0 \\ 1 \\ 1 \\ 0 \end{array} \right) \\
 \vert \psi_- \rangle =   {\rm CNOT}\; \mathbf{H}\otimes \mathbf{I}\; \vert 10\rangle &=& 
    \frac{1}{\sqrt{2}} \left( \begin{array}{cccc} 1 \\ 0 \\ 0 \\ -1 \end{array} \right) \\
   \vert \phi_- \rangle =  {\rm CNOT}\; \mathbf{H}\otimes \mathbf{I}\; \vert 11\rangle &=& 
    \frac{1}{\sqrt{2}} \left( \begin{array}{cccc} 0 \\ 1 \\ -1 \\ 0 \end{array} \right) .
\end{eqnarray}
%
and $\mathbf{H}$ is the 
Hadamard operator, 
\begin{equation}
    \boxed{\rm\mathbf{H}} \equiv \frac{1}{\sqrt{2}} \begin{pmatrix}  1 & 1 \\ 1 & -1 \end{pmatrix}.
    \label{e:H}
\end{equation}. 


All four entangled states $-$ commonly referred to as $\vert \phi_\pm\rangle$ and $\vert \psi_\pm\rangle$ as shown above $-$ are central to the construction of a quantum interferometer. There is another useful operation specific to the Bell states, the controlled Z-gate or $\boxed{\mathbf{CZ}}$, that simply flips the sign of $\vert 11\rangle$ and leaves the rest alone.

Other important gates are even simpler and include:
\begin{eqnarray}
{\rm Identity\;operator}\;\;\boxed{\rm\mathbf{I}} &\equiv& \begin{pmatrix}  1 & 0 \\ 0 & 1 \end{pmatrix} \\
{\rm NOT\; or\; X\text{-}gate\; operator}\;\;\boxed{\rm\mathbf{X}} &\equiv&  \begin{pmatrix}  0 & 1 \\ 1 & 0
\end{pmatrix}\;\; \equiv\;\; \boxed{\mathbf{NOT}} \\
{\rm Y\text{-}gate\; operator}\;\;\boxed{\rm\mathbf{Y}} &\equiv&  \begin{pmatrix}  0 & -i \\ i & 0
\end{pmatrix} \\
{\rm Z\text{-}gate\;operator}\;\;\boxed{\rm\mathbf{Z}} &\equiv&  \begin{pmatrix}  1 & 0 \\ 0 & -1
\end{pmatrix}
\end{eqnarray}

\paragraph{Quantum optical analogues}

The quantum operators acting on the qubits in the previous sections, i.e., CNOT and Hadamard, have practical optical analogues.  Indeed, the experimental apparatus for carrying out these remarkable operations is readily available to University research labs\cite{deh02}. 

{\bf Phase retarder}.
Consider two photons, one that propagates horizontally ($\vert H\rangle$) corresponding to logical bit $\vert 0\rangle$, and another propagating vertically ($\vert V\rangle$) associated with logical bit $\vert 1\rangle$. These states are equivalent but the former are often associated with experiment, and the latter with mathematical notation. We can use these states to define/generate single and double state qubits. For example, we can produce a ``superposition by polarisation'' by passing $\vert 0\rangle$ or $\vert 1\rangle$ through a birefringent crystal known as a half-wave plate (HWP):
\begin{center}
\includegraphics[scale=0.55]{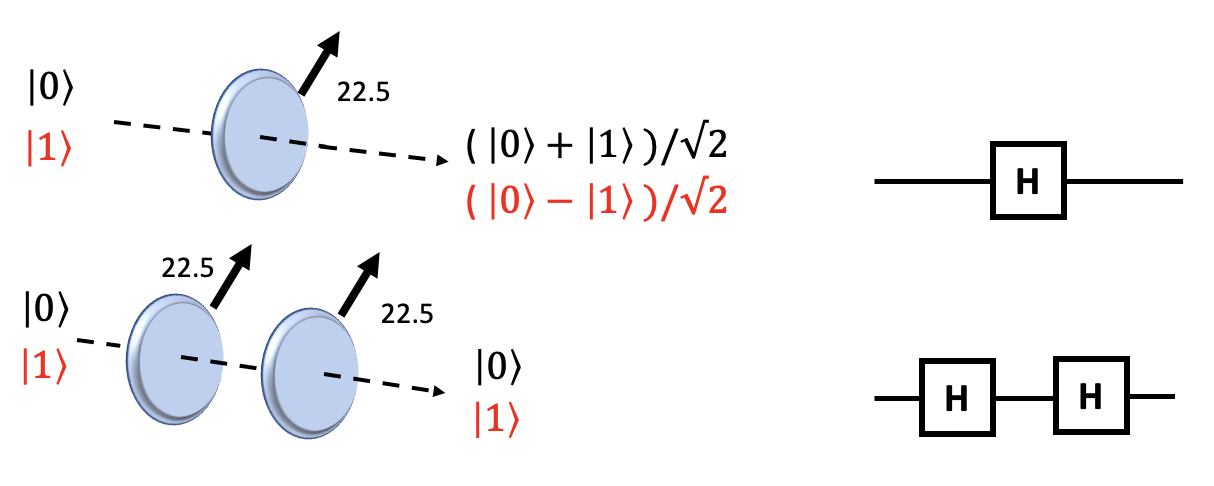}
\end{center}
The use of HWPs at an angle $\theta=\pi/8$ to the optic axis (black arrow) is common
although a polariser at twice the angle also works if the polarised source is aligned. The HWP rotates {\it linearly polarised} light at an angle $-\theta$ to the optic axis and therefore rotates the polarisation vector by $+2\theta$.

The logic gate symbols (Hadamard gates) are shown on the RHS. The black and red input photonic states map to the black and red superpositions, respectively. Note that two HWPs in a row transform back to the original polarisation state. 

For completeness, the output Bell states above are two of the most famous qubits in quantum computing, and they have their own widely used symbols, i.e. 
\begin{eqnarray}
    \vert +\rangle &=& \mathbf{H}\vert 0\rangle = \frac{1}{\sqrt{2}} (\vert 0\rangle + \vert 1\rangle)\;\; \equiv \;\;
    \frac{1}{\sqrt{2}} \begin{pmatrix} 1 & 0 \\ 0 & 1 \end{pmatrix} \\
     \vert -\rangle &=& \mathbf{H}\vert 1\rangle = \frac{1}{\sqrt{2}} (\vert 0\rangle - \vert 1\rangle)\;\; \equiv \;\;
    \frac{1}{\sqrt{2}} \begin{pmatrix} 1 & 0 \\ 0 & -1 \end{pmatrix}
\end{eqnarray}
It follows trivially that $\mathbf{HH}\vert 0\rangle=\vert 0\rangle$ and $\mathbf{HH}\vert 1\rangle=\vert 1\rangle$. The Hadamard operator must be modified for beams that are split, rather than in-line, as we show in the next section.
Note that these optical circuits allow us to perform operations on states that cannot be ``read'' without destroying them.

{\bf Spatial beam splitters and the Mach-Zehnder interferometer}.  We can also produce a ``superposition by path'' 
using a non-polarising spatial beam splitter (SBS) or, more commonly, a polarising beam splitter (PBS) that has an extra degree of freedom. We start with the 50:50 SBS fed with individual photons one at a time, either along beam axis ``a'' or ``b'':
\smallskip
\begin{center}
\includegraphics[scale=0.5]{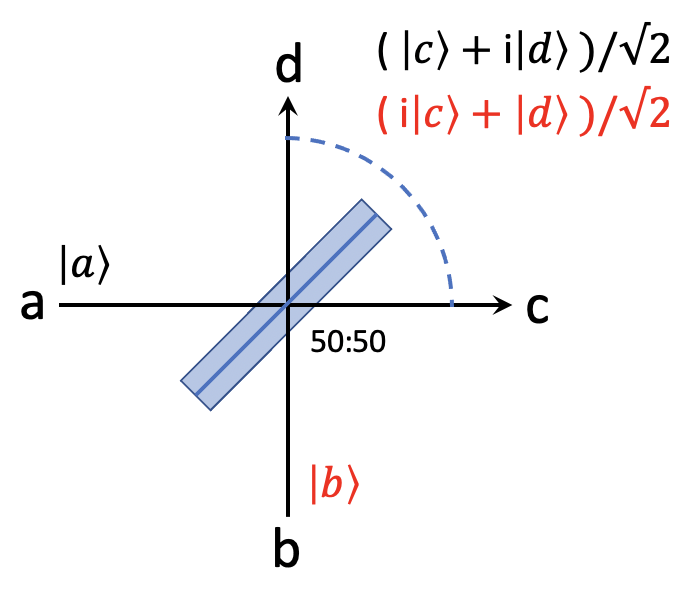}
\end{center}
Once again, black ($\vert a\rangle$) and red ($\vert b\rangle$) input photons map to the black and red superpositions indicated here, respectively. The single qubit states in red and black are made possible by the superposition of two output paths from an input beam, indicated by the dashed arc. The $i$ indicates the phase shift ($\frac{\pi}{2}$) of the reflected beam. (Quantum systems often use both input beams ``a'' and ``b'' simultaneously in order to minimise coupling an empty input beam to the vacuum field; the (zeropoint) noise fluctuations are easily seen in modern experiments, e.g. when working with entangled photons. We discuss simultaneous input photons below.)

We can now build more complex functions, starting with the {\it Mach-Zehnder interferometer} (MZI) comprising two 50:50 SBS and two mirrors (M), and a phase retarder/polariser (P), such that we get one of the primary building blocks of quantum devices:
\begin{center}
\includegraphics[scale=0.5]{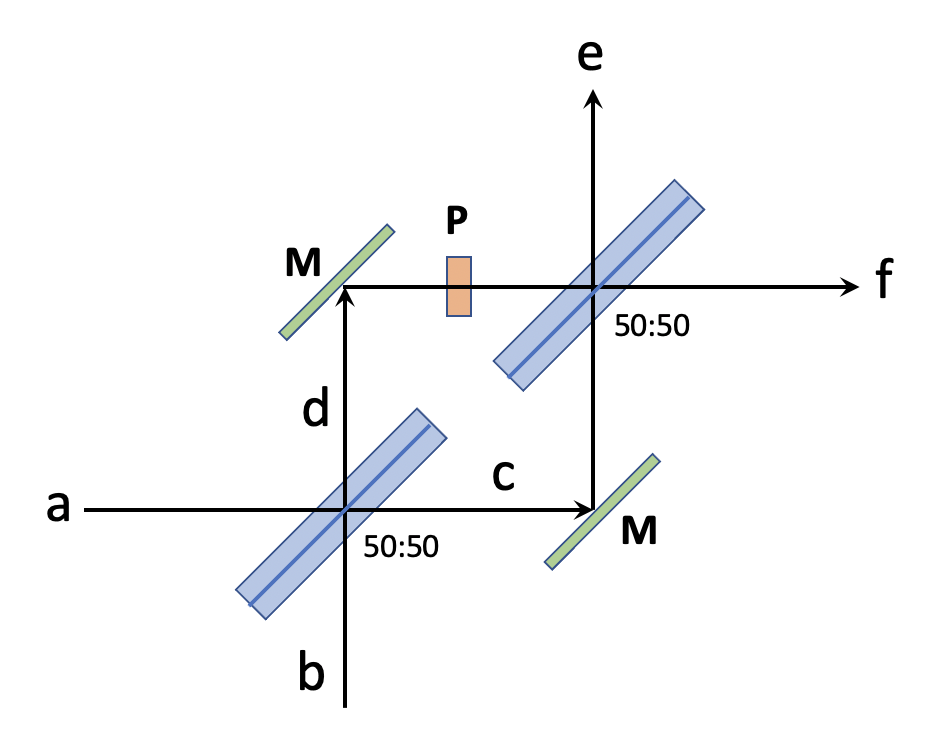}
\end{center}
The beams ``c'' and ``d'' are transformed in the following way:
\begin{eqnarray}
    \vert c\rangle &\rightarrow& \frac{1}{\sqrt{2}}(\vert e\rangle + i \vert f\rangle ) \\
     \vert d\rangle &\rightarrow& \frac{ e^{i\varphi}}{\sqrt{2}}(i\vert e\rangle +  \vert f\rangle ) 
\end{eqnarray}
where the phase shift $\varphi$ is imposed by P. Thus, it follows trivially that
\begin{equation}
    \vert a\rangle \rightarrow i e^{i\varphi/2} \left( -\sin\frac{\varphi}{2}\vert e\rangle + \cos\frac{\varphi}{2}\vert f\rangle \right)
\end{equation}
After converting amplitudes to probabilities, we see that the probabilities of photons emerging at the outputs are
\begin{eqnarray}
    P(\vert e\rangle) &=& \sin^2 \frac{\varphi}{2} = \frac{1}{2}(1-\cos\varphi) \\
     P(\vert f\rangle) &=& \cos^2 \frac{\varphi}{2} = \frac{1}{2}(1+\cos\varphi) 
\end{eqnarray}
Note that we can choose the phase shift such that the entire MZI simply behaves as a monolithic beamsplitter or that the signal is all in one or other channel. This last case is interesting because it has a profound meaning $-$ the photon {\it always} travel along both paths (as for Young's slits) because it is interference that destroys one of the paths. We can verify that by moving the detector further away along the ``e'' or ``f'' axis and see the counts maximize and minimize because of the changing time delay\cite{hug21}.

{\bf Polarising Beam Splitters}.
In the last section, the spatial beam splitter operates on logic values $\vert 0\rangle$ and $\vert 1\rangle$ equivalent to a matrix rotation, i.e.
\begin{equation}
    B(t,r) = 
    \left( \begin{array}{cc}
    t & ir \\
    ir & t \end{array} \right)
\longrightarrow \boxed{\mathbf{H}} = 
    \frac{1}{\sqrt{2}}\left( \begin{array}{cc}
    1 & i \\
    i & 1 \end{array} \right)
\end{equation} 
where $t$ and $r$ are the transmission and reflectance coefficients of the beamsplitter coating. We obtain the non-polarising SBS after we specify how the coatings are to respond to $H$ and $V$ polarisation, i.e. $r_V=r_H$ and $t_V=t_H$. This matrix operation is equivalent to the spatial Hadamard operation.

A polarising beamsplitter specifies different coatings, i.e. $r_V=1$, $t_H=1$, $r_H=0$ and $t_V=0$. Let us look at the case specific to Bell state tests. For a superposition on input across beams ``a'' and ``b'', we must consider
\begin{equation}
    \begin{pmatrix}
        a_H \\
        b_H \\
        a_V \\
        b_V
    \end{pmatrix}
\end{equation}
The $H$ component of beam ``a'' is orthogonal to the splitting surface and so is transmitted; the $V$ component is parallel and so is reflected. Thus
\begin{center}
\includegraphics[scale=0.5]{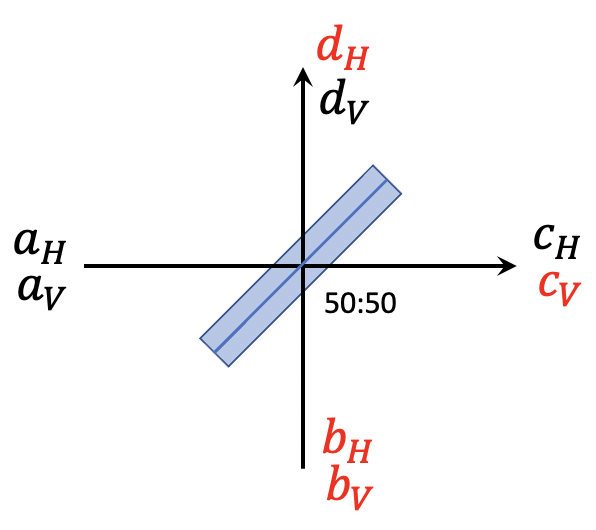}
\hspace{1cm}
\includegraphics[scale=0.4]{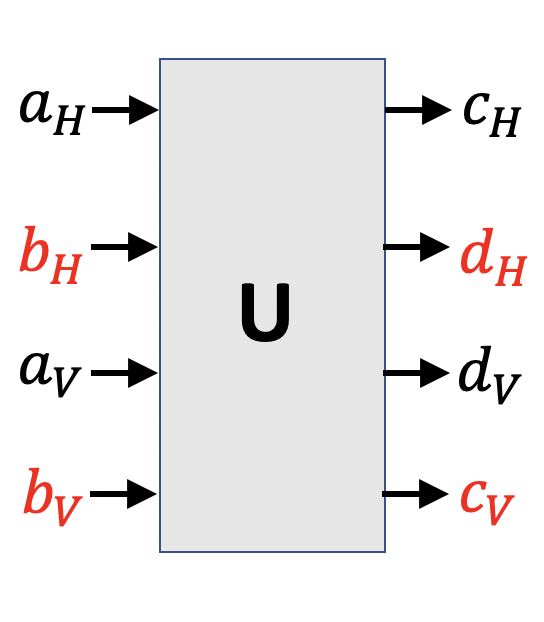}
\end{center}
The $1\times 4$ column matrix resembles the two-qubit state vector, which we return to below. It is therefore not surprising that PBS devices are commonly employed in two-qubit state operations and, in fact, in the construction of two-qubit gates.

The matrix operation (unitary operator) is
\begin{equation}
 \boxed{\mathbf{U}} = \left( \begin{array}{cccc}
    t_H & ir_H & 0 & 0 \\
    ir_H & t_H & 0 & 0 \\
    0 & 0 & t_V & ir_V \\
    0 & 0 & ir_V & t_V 
    \end{array} \right) = 
    \left( \begin{array}{cccc}
    1 & 0 & 0 & 0 \\
    0 & 1 & 0 & 0 \\
    0 & 0 & 0 & 1 \\
    0 & 0 & 1 & 0 
    \end{array} \right) \equiv \boxed{\mathbf{CNOT}}
\end{equation}
The $\boxed{\mathbf{U}}$ operator has a well-defined relationship between the optical layout and the operator definition. This is identical to the so-called controlled NOT gate or $\boxed{\mathbf{CNOT}}$ gate that lies at the heart of quantum computers (see section~\ref{sec:entangle}).

\paragraph{Natural sources for two-qubit entangled photons}

Entanglement is a strange and remarkable property of nature\footnote{Cosmologists tell us that everything in the Universe is entangled with an unknown fraction of entangled pairs having a partner beyond the observable horizon. This is also the basis for Hawking radiation from a black hole.}, for which there are different paths for this to occur naturally. However, natural sources of entangled photons tend to be far too impractical or inefficient for our applications. For example, Aspect's 1982 experiment\cite{asp82} (and earlier attempts) used a cascade transition. In these experiments, electrons within calcium atoms were put into a highly-excited energy level (Rydberg state) for which the electron is unable (forbidden) to return to the ground state directly by emitting a single photon. Instead, the electrons decay via a short-lived intermediate state, releasing two photons with different energies within $\sim 1$ ns of each other. The biphotons have correlated polarisations, and are radiated in random but opposite directions. These sources are too impractical for modern use.


The 1967 discovery of spontaneous parametric down-conversion (SPDC) eventually revolutionized quantum optics and photonics and, arguably, enabled much of the subsequent quantum revolution\cite{har67}. In essence, a 405nm diode laser beam is fired into a non-linear ($\chi^{(2)}$) optical crystal (typically beta barium borate, or BBO). We align the (mostly polarised) incoming laser beam with a 45$^\circ$ polariser, which is a superposition of $H$ and $V$ polarisation, to clean the light. 
The individual high energy photons leaving the polariser are in a superposition of $H$ and $V$.
These interact with the quantum vacuum and the crystal to produce biphoton pairs with wavelength 810 nm. These propagate in the direction of the original beam, but on opposite sides of a cone with opening angle of a few degrees:
\begin{center}
\includegraphics[scale=0.4]{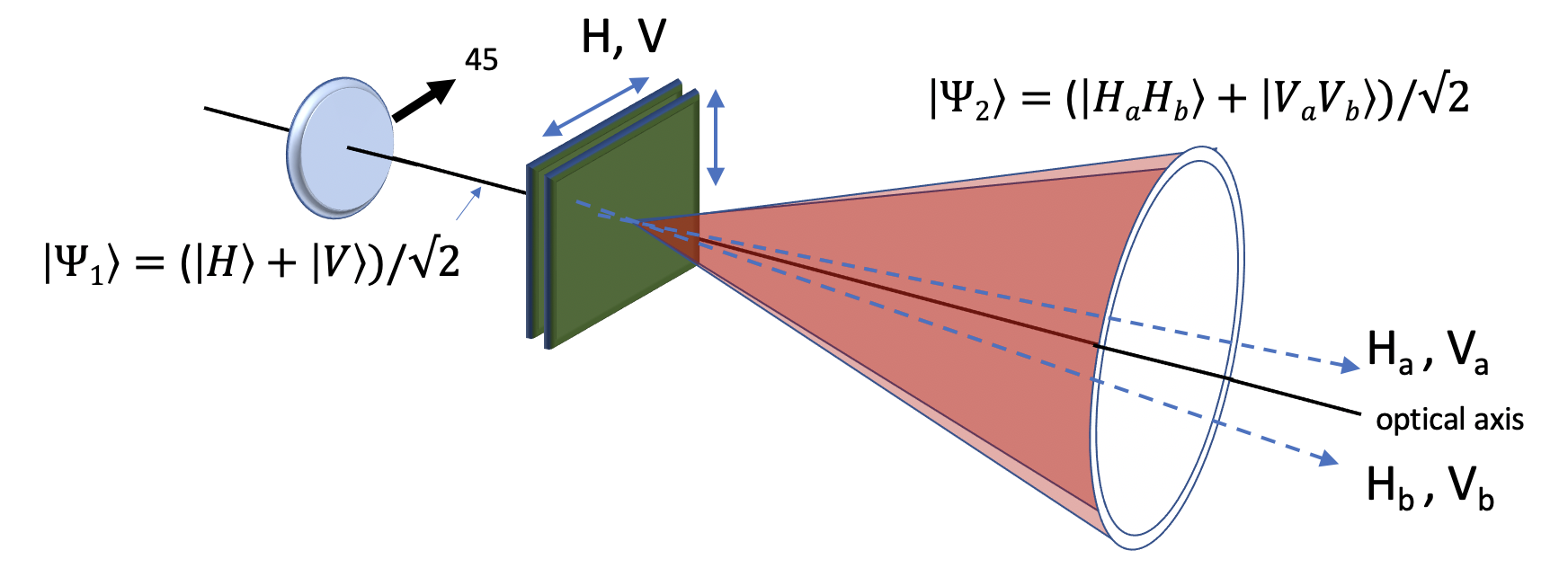}
\end{center}
If the crystal is cut correctly and bonded as two layers, the biphotons will have identical polarisation and be entangled. This is how. We orient one of the BBO crystals to interact with horizontally polarised ($H$) light, the other with vertically polarised ($V$) light. For an $H$ incoming photon, on occasion, it produces two IR photons with $V$ polarisation, and vice versa. Now each incoming photon has an
equal probability of interacting with either of the crystals (but not both). For two photons produced in 
the same crystal (biphotons) propagating on the surface of a cone,
\begin{equation}
    \vert \psi_2 \rangle = \frac{1}{\sqrt{2}} \left( \vert H\rangle_a \vert H\rangle_b + \vert V\rangle_a \vert V\rangle_b  \right) \equiv
    \frac{1}{\sqrt{2}} \left( \vert HH\rangle_{ab} + \vert VV\rangle_{ab}  \right)
\end{equation}
for which $a$ and $b$ refer to IR photons on opposite sides of the cone. Thus, our single qubit state has produced two pairs of entangled photons along two beams that are in a superposition.

For the entangled two-qubit state $\vert \psi_2 \rangle$, for any polarisation angle $\alpha$, photon $a$ is detected half the time in state $V$, 
and half the time in state $H$. 
The same holds true for photon $b$
for some other polarisation angle $\beta$. In this entangled state, the photons have no intrinsic prior polarisation that we can know about. Indeed, the polarisation is imposed {\it randomly} by the measurement -- there are no hidden variables. 
Remarkably, if we set $\alpha=\beta$, once the polarisation for photon $a$ is known, the other polarisation for photon $b$ is determined, i.e. 
identical to photon $a$. These form an entangled pair that are maximally entangled when $\alpha=\beta=\pi/4$.
The conditional probabilities are $P(V_b^\alpha\vert V_a^\alpha )=P(H_b^\alpha\vert H_a^\alpha )=1$ for all $\alpha$.
In a practical system, the beam on each side of the cone passes through an 810 nm narrowband filter before entering a collimator that focusses to an imager/coincidence monitor.
The general formula for different polariser orientations is given by $P(\vert V_a^\alpha\rangle\; \vert\; \vert V_b^\beta\rangle) = \cos^2(\alpha-\beta)$ whose maximum value occurs at $\alpha=\beta.$






\section{Practical applications}
\label{sec:pracapp}

\subsection{Hexabundle spectrograph}

The simplest application of astrophotonics is the use of fibres to transport light from one place to another, most often from a focal plane to the input slit of a spectrograph as for multi-object spectroscopy\cite{hil80}.  Despite being very simple from a photonics functionality viewpoint the power of this technique should not be underestimated having transformed our understanding of the large scale structure of the Universe, galaxy evolution, and the formation history of the Milky Way.

More recently these ideas have been extended to multi-object spatially resolved spectroscopy through the use of fused fibre bundles.  In this way each fibre in the bundle can be used to sample only one part of a spatially contiguous region, allowing individual spectra from across the object to be recorded.

Non-photonic solutions exist for integral field spectroscopy as well.  Image slicers use stacks of fanned mirrors to divide the field and reconstruct the image into a slit.  These are very efficient, but also complex, and challenging to incorporate into a multi-object system, although this has been done for the KMOS instrument\cite{sha13}.  Another solution is to use microlens arrays to sample the image, but this leads to a rather inefficient use of the detector, and non-uniform spectral coverage.

Photonic solutions using fibres offer the flexibility to easily reposition the bundles, making them very well suited for multi-object integral field spectroscopy, but in addition, they allow uniform wavelength coverage, good use of the detector, and coupling from a fast telescope beam (modern telescopes are made with intrinsically fast beams to enable compact and stiff structures).

A notable example of this technology is the hexabundle\cite{bland11,bry14}.  Here the cladding of the fibres is stripped down to a thin layer ($\sim 5$~$\mu$m), and then lightly fused into a bundle over $\sim 2$~cm, such that the cores of the fibres are not deformed, which otherwise would lead to unacceptable focal ratio degradation\cite{bland11}.  This allows efficient sampling ($> 70$ per cent filling factor), while minimising cross-talk and FRD.  Hexabundles have been successfully used in the SAMI Galaxy Survey\cite{bry15,cro12}, a survey of 3400 galaxies each observed with a 61-core hexabundle, with 1.6 arcsec sampling.

\subsection{Microspectrograph}
\label{sec:microspect}

As telescopes get larger so must the instruments installed on them, since the size of a spot at the focal plane is given by,
\begin{equation}
    {\rm spot\ size} = \frac{\Gamma f D}{206265},
\end{equation}
where $\Gamma$ is the size of the object in arcsec, $f$ is the telescope focal ratio, and $D$ is the diameter of the telescope.  Therefore as the diameters of new telescopes increase, classical instruments must become larger or faster, or both.

This scaling can be broken by using replicated, modular limited instruments, each of which samples a segment of the image.  If these modular instruments are diffraction limited and fed by single mode fibre, then they can be made as small as physically possible.

One way to achieve this is to use adaptive optics to achieve diffraction limited resolution.  However, for efficient coupling, a high Strehl ratio (low wavefront error) is required, which limits this technique to specialised AO systems.  Currently these are limited to wavelengths $>1$~$\mu$m, and furthermore have limited sky-coverage, since a bright natural guide star is required for tip-tilt correction.  Coupling into SMF has been achieved with efficiencies of $>40$\%\ using SCExAO on the Subaru Telescope\cite{jov17}, which could be increased up to 67\%, approaching the theoretical limit of 80\%\cite{sha88}.

An alternative method is to use photonic lanterns to capture  a seeing-limited image with a multimode fibre, and then to convert this into an array of single mode fibres\cite{leo05,bir15}.  However, in poor seeing, or for large telescopes (with large \'{e}tendue) a very high number of single mode fibres may be required, see equation~\ref{e:modesee} (essentially, a larger image will require a larger fibre, which carries more modes).  This could be alleviated with a `divide-and-conquer' scheme in which bundles of multi-core fibres are first split into individual multimode fibres, and are then later split into SMF\cite{leo17}.  A combination of lower Strehl AO with low mode count photonic lanterns could provide a good compromise between these techniques, increasing sky coverage and wavelength range without the need for either extreme AO or very high numbers of modes.

Proof-of-concept prototype microspectrographs have been tested on-sky.  Using a $1 \times 7$ photonic lantern to feed a diffraction limited spectrograph made from commercial off-the-shelf components  Betters et al.\cite{bet13}
measured a solar spectrum in the near-infrared with a throughput of 60~\%\ and a resolving power of $\approx 30,000$.  Due to the Gaussian input beam profile and the diffraction limited optics, this device showed remarkably little scattering, with a line spread function close to the theoretical limit.  Subsequent developments used a $1 \times 19$ photonic lantern to feed a diffraction limited \'{e}chelle spectrograph with a resolving power of $\approx 60,000$, which was tested at the UK Schmidt Telescope to obtain a spectrum of $\alpha$-Centauri over the wavelength range 655 -- 770~nm\cite{bet14}.

AWGs (section~\ref{sec:disperse}) have also been tested on-sky.  A prototype was tested on the AAT, using IRIS2 as a cross-disperser, with a free-spectral range of 57~nm and a resolving power of $\approx 2100$ at a wavelength of 1500~nm\cite{cve09}, which was used to make the first astronomical observations with an integrated photonic spectrograph, measuring the CO absorption bands in $\pi$~1~Gru\cite{cve12a}.  Further refinement of AWGs designed specifically for astronomy has resulted in devices capable of high resolving power ($R\approx 12,000$) with low free spectral range ($\approx 18$~nm)\cite{gat21,gat22}.

Photonic \'{e}chelle gratings are an interesting, and still relatively unexplored technology\cite{wat96}.   These devices are conceptually very similar to a standard \'{e}chelle reflection grating, but the grating is written into a slab waveguide and thus the number of reflecting facets can be very much larger than the number of individual waveguides in an AWG.  Therefore the resolving power $R=m N$  can be correspondingly larger.  Furthermore, they can be used directly with few-moded fibres without loss of performance relevant to astronomical applications\cite{bland06}.  Ironically, despite being one of the very first integrated photonic devices suggested for astronomy\cite{wat95}, and being seemingly more suited and more easily adapted to astronomy, they have attracted less attention because they have proved harder to  adapt to the needs of telecommunications\cite{pat14}, especially as regards polarisation dependence and birefringence and very strict requirements on cross-talk\cite{bland06}, and therefore manufacturing methods lag behind those of AWGs.  A notable exception is the recent work by Stoll et al.\cite{sto19} who have investigated various designs for PEGs for astronomical spectroscopy.

Lippmann spectrographs (section~\ref{sec:disperse}) have been developed for astronomy\cite{con95,lec07,bli17}, but have found more use in other areas.
The primary difference between Lippmann spectroscopy and the other techniques described above is that in a conventional diffraction grating spectrograph or an AWG the light is dispersed such that each pixel sees a different wavelength.  On the other hand, in a Lippmann spectrograph all pixels see a contribution from all wavelengths.  This can be either advantageous or disadvantageous depending on the source of noise, but suffers from the same signal-to-noise disadvantages as a Fourier Transform Spectrograph for astronomical sources with continuous spectra.

\subsection{Photonic wavefront sensor}

A wavefront sensor typically consists of an array of subapertures that sample the incoming wavefront. Each subaperture contains a small lens or micro-optical element, which focuses the light onto a detector or a pixel in a camera. By analyzing the intensity distribution of the focused light, the wavefront sensor can determine the phase and amplitude information of the incident wavefront. There are different types of photonic wavefront sensors, including Shack-Hartmann sensors, shearing interferometers, and pyramid wavefront sensors. Each type has its own advantages and applications, but they all essentially work by sampling the wavefront and analyzing the resulting light distribution.

This same action has now been demonstrated in a photonic lantern\cite{nor20}.  The coupled power into each SMF output will depend on the input wavefront, as well as the wavelength of light and the geometry of the photonic lantern.  Assuming these latter variables to be fixed, the measured power in each output SMF can then be used to reconstruct the input wavefront.  Furthermore, in principle the corrected image can be reconstructed directly from these measurements, without the need to perform the corrections in real time.

\subsection{Sky suppression spectrograph}
\label{sec:ohsupp}

The night sky is very bright at near-infrared wavelengths due to emission from OH radicals at $\approx 88$~km altitude.  The resulting forest of rovibrational lines is difficult to subtract accurately because it varies rapidly, both temporally and spatially\cite{con96,ell08}.   This is problematic for many areas of astrophysics which require deep near-infrared spectroscopy, e.g.\ the study of highly redshifted galaxies, low mass stars, and star-formation in our own Galaxy. 

Traditional techniques to subtract or filter this emission are limited by the real-world performance of spectrographs, in particular stability and scattering\cite{ell08}.  A photonic solution has been developed and demonstrated on-sky using fibre Bragg gratings\cite{bland04,bland08,bland11b}, see section~\ref{sec:filter}.  

\begin{figure}[t]
    \centering
    \includegraphics[scale=0.45]{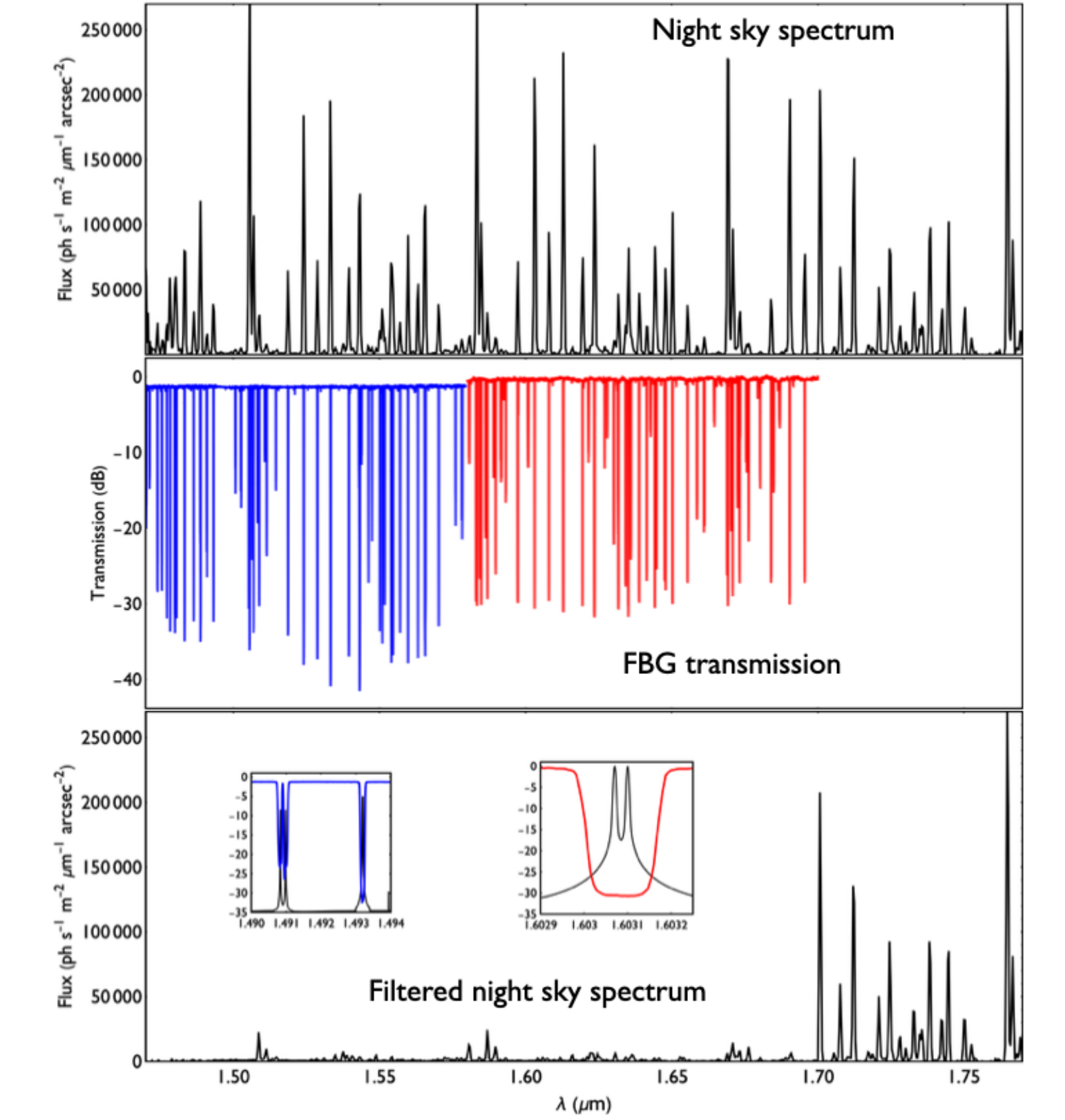}
    \caption{The measured response of the GNOSIS fibre Bragg gratings\cite{tri13a} are shown in the middle panel, compared to the night sky spectrum (top panel), and the resulting suppressed sky spectrum (bottom panel)\cite{ell12a}.}
    \label{fig:fbg}
\end{figure}

OH suppression with fibre Bragg gratings has been demonstrated in two on-sky experiments, GNOSIS\cite{tri13a,ell12a}, and PRAXIS\cite{ell20}.  The latter demonstrated an overall peak end-to-end efficiency, from atmospher to detector, of 18 per cent, whilst reducing the integrated night sky background by a factor of 9, and suppressing individual OH lines by factors of up to 10,000.  However, PRAXIS suffered from high thermal background (\emph{not} due to the photonic components), and so far FBGs have not been used for scientific observations despite their obvious benefit.

Alternative solutions to single mode FBGs are being pursued, primarily with the aim of allowing easier scaling to higher numbers of modes, and therefore large fields-of-view.  These include attempts to write FBGs into fibre more efficiently, for example using multicore fibre\cite{lin14}, or using direct-write techniques into fibre\cite{goe18}.  Alternatively Bragg gratings can written directly into waveguides incorporated with photonic lanterns using femtosecond laser inscription\cite{spa14b}. Lithographic techniques have also been explored using either complex waveguides\cite{zhu16}, i.e.\ waveguides with variation in width to alter the effective index, or microring-resonators\cite{ell17,liu21}. 

\subsection{Photonic nulling}

Nulling interferometry is an important and promising technique for the direct detection and analysis of exoplanets.  In this technique light from two apertures (e.g.\ two telescopes) is interfered   to obtain the coherent superposition of both sources, from which the spatial geometry of the source can be obtained.  By adding a $\pi$ phase shift to one arm, the superposition will be out of phase on axis, leading to a deep null, and effectively cancelling the light from the host star.  The off-axis light from the planet will not be nulled.

\begin{figure}
    \centering
    \includegraphics[scale=0.37]{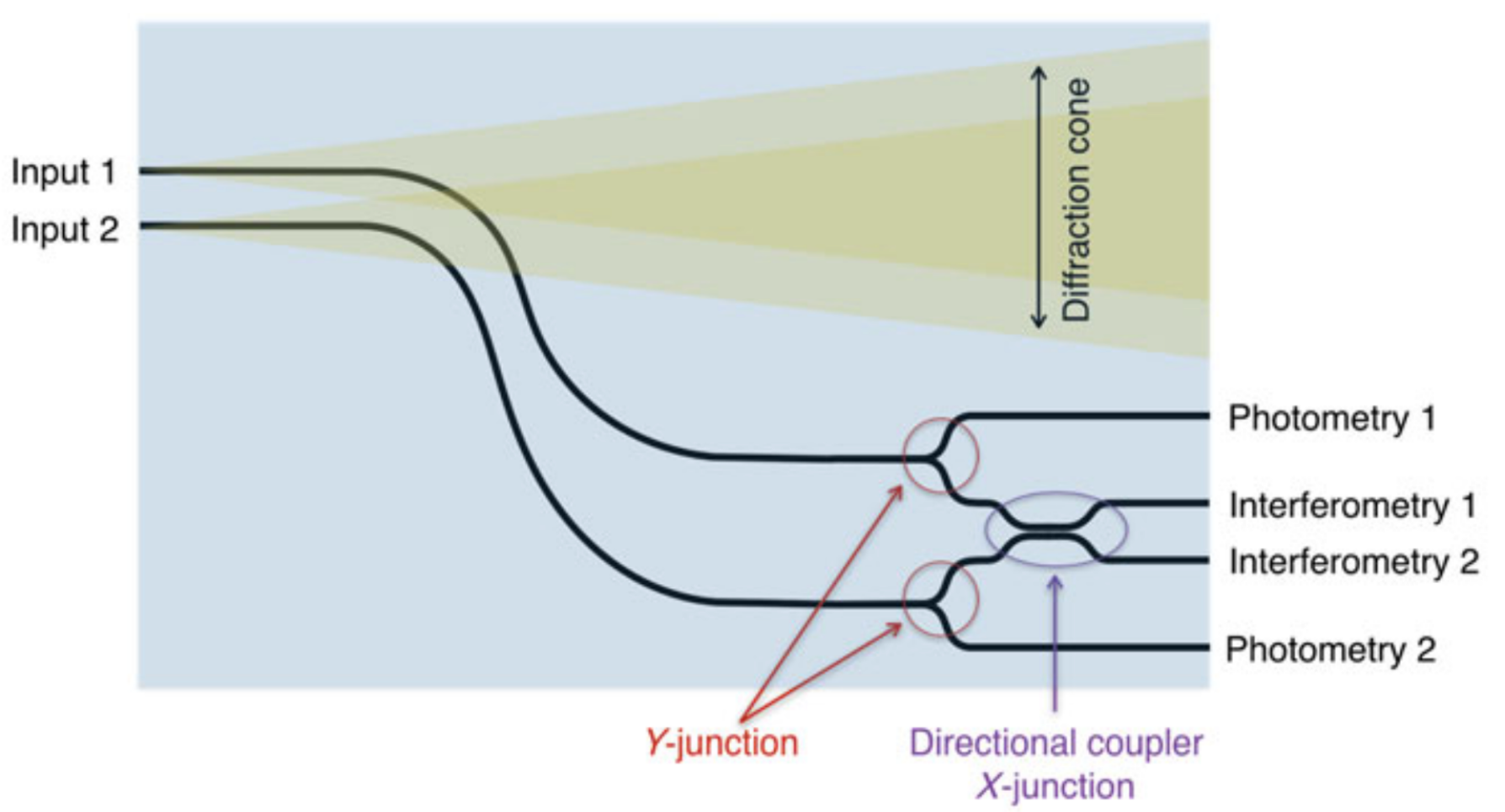}
    \caption{Schematic diagram of a photonic nulling interferometer\cite{lag21}.
    }
    \label{fig:nuller}
\end{figure}

This can be accomplished using integrated photonics, for example as shown in Figure~\ref{fig:nuller}.  Here light from two apertures is injected into separate waveguides, which are then interfered in an evanescent coupler, which incorporates a $\pi$ phase shift in one arm.  Additionally, some light from each input waveguide is split off for photometric calibration.  For efficient performance and high-contrast nulling AO injection is required; using SCExAO on the Subaru telescope to inject into a photonic nuller, null depths of $10^{-3}$ with a precision of $10^{-4}$ have been achieved\cite{nor20,mar21}.

The above scheme is limited to narrow passbands since the evanescent coupler is strongly wavelength dependent (light of different wavelength will not remain in phase as it propagates down the waveguides due to the different mode profiles and different effective index).    However, recent advances using photonic tri-couplers are able to achieve achromatic nulling, as well as incorporating fringe-tracking.   Here light is injected into 2 out of three waveguides, and with a $\pi$ phase shift between them.  If the coupler is symmetric, such that the coupling between each of the two input waveguides and the third waveguide is identical, then no light can couple into the third waveguide, regardless of wavelength, since the coupled light from each waveguide is out-of-phase.  Moreover, if the coupling between all three waveguides is identical, as for waveguides arranged as the vertices of an equilateral triangle, then there will be an equal splitting ratio between the two outer channels, allowing the input phase delay to be recovered for fringe-tracking\cite{mar21b,kli22}.

\begin{figure}
    \centering
    \subfigure[]{
    \includegraphics[scale=0.35]{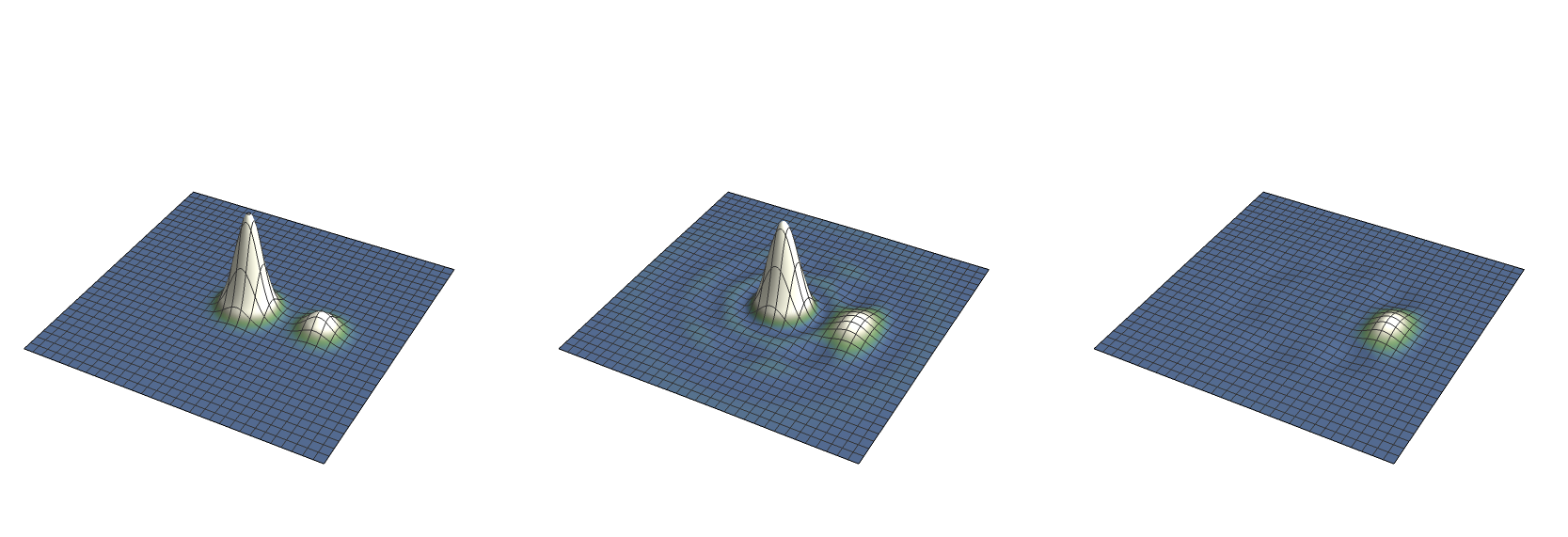}
    }
    \subfigure[]{
    \includegraphics[scale=0.35]{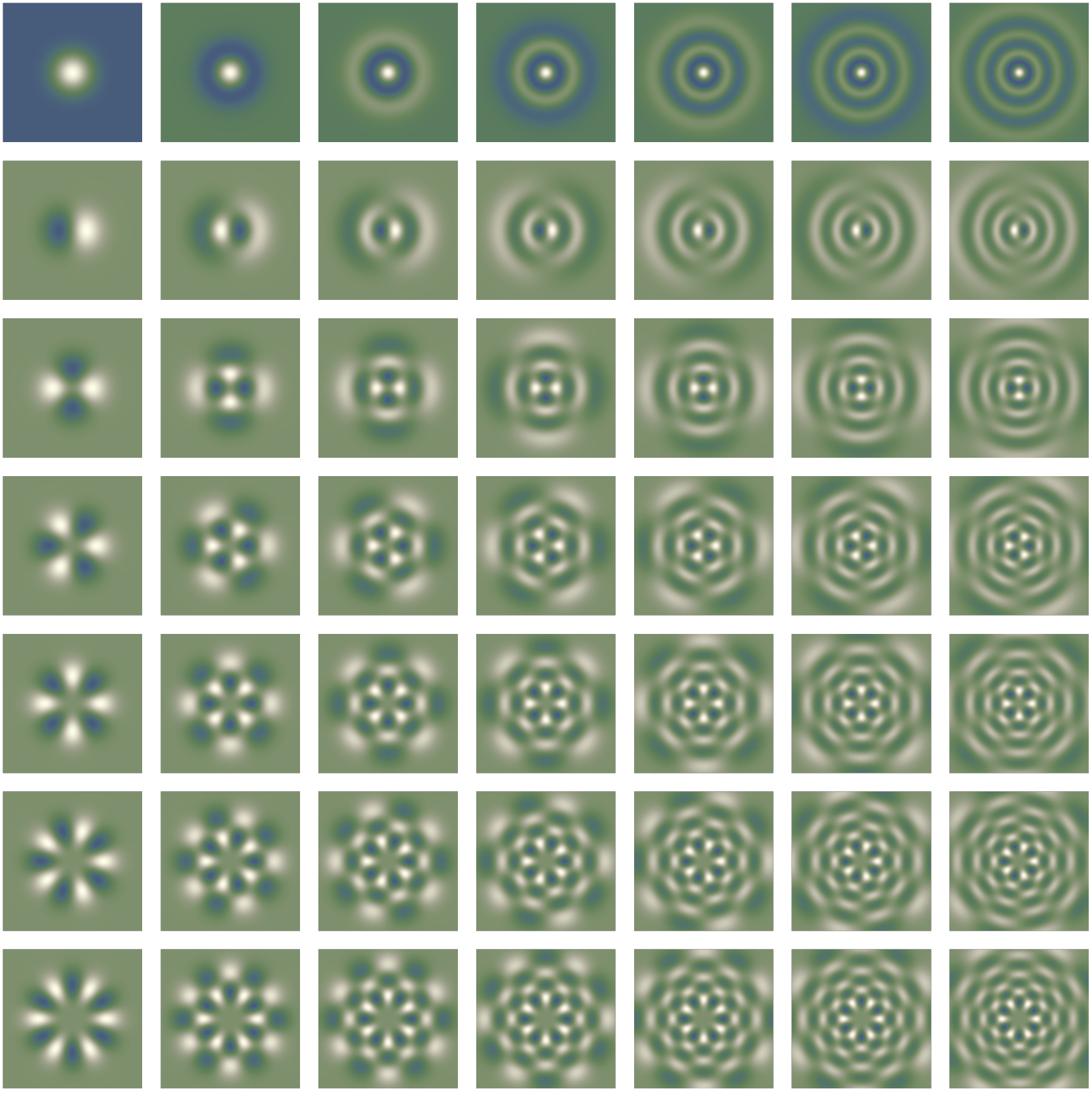}
    }
    \subfigure[]{
    \includegraphics[scale=0.35]{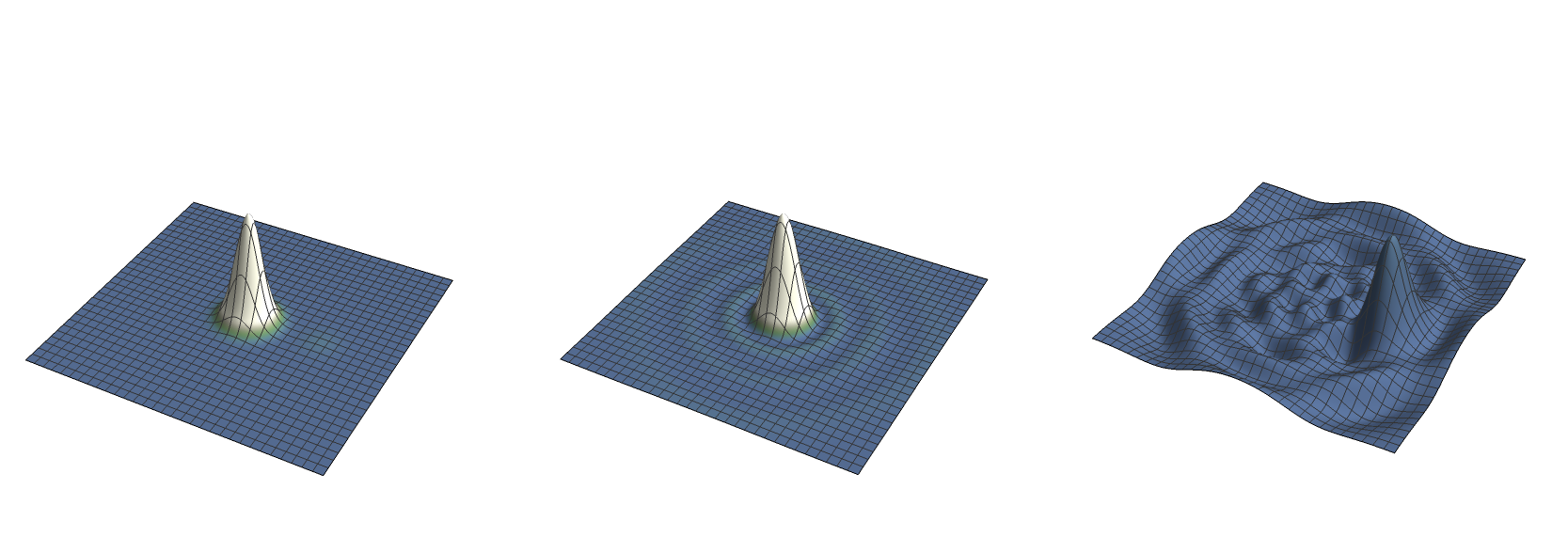}
    }
    \caption{A simulated image of a star and planet (top left), the reconstuction from the fitted Laguerre Gaussian modes (top middle), and the reconstruction from without the $l=0$ modes (top right).  The middle panel shows the first 49 LG modes, from which the image was reconstrcuted.  The bottom panel shows a second example with a fainter planet, which is not apparent until the on-axis modes have been removed. }
    \label{fig:lgmodes}
\end{figure}

A different solution is to perform coronagraphy by blocking or removing the on-axis starlight from an image.  This can be done by means of a physical mask, but scattered light from the central star will still affect the neighbouring regions.  A photonic solution is offered by means of decomposing an image into its Laguerre-Gaussian modes, and then filtering the on-axis axisymmetric modes, before reconstructing the image, in effect performing a type of photonic coronagraphy.  This has been demonstrated in the lab, using a spatial light modulator to decompose the image into its LG modes with potential benefits for throughput, inner-working angle, aberration tolerance\cite{fon19,car20,car20b}.  Simulated examples are shown in Figure~\ref{fig:lgmodes}.

A similar method uses a phase mask to decompose an image into its Fourier components, and then to filter the spatially symmetric components containing the on-axis starlight with a Lyot stop\cite{foo05}.  This advanced method is now routinely used in several instruments\cite{swa08}, with future advances capable of achieving contrasts better of $\approx 10^{-8}$\cite{llo20}.

\subsection{Optical interferometer}
\label{sec:interferometer}


All forms of astronomical interferometry work on the van Cittert-Zernike principle. 
The source of photons is always highly incoherent but the distances are so vast that the observed wavefront is coherent on arrival, except for the Earth's incoherent atmosphere which imposes a short coherence time (and length) on the observations. The theorem states that the complex degree of coherence at the detector (modulo the atmosphere) is the Fourier transform of the intensity distribution at the source. Typically, one optical path is varied so that the fringe visibility is obtained as a function of that path-length.

Thus, widely spaced telescopes observing the same faint star can be used to generate phase-preserved interference fringes in the limit of discrete photon events. But in order to do this, we must maintain mutual coherence at all telescopes with respect to the same incoming coherent wavefront. For optical/IR photons, this is tricky. Radio interferometers achieve this by measuring a continuous stream of fluctuating voltages (not photons) and mixing the signals at different telescopes with an identical local oscillator source. These machines have the advantage of measuring amplitude and phase directly and, moreover, the electronic capability to mix this signal with a common reference frame, e.g.\ a local oscillator.

The application of astrophotonics to interferometry can be traced back to  the suggestion of using single-mode fibres to transport the light from telescopes to the beam combiners by Froehly (1981) \cite{fro81}. Today the main area of application is in the beam combination itself.  The combiner works via the principle of evanescent coupling described in section~\ref{sec:combine}.

The first astrophotonic beam combiners used single-mode fibres to combine the beams from two telescopes, but these have now been superseded by integrated photonic beam combiners which are more stable and can be more easily extended to multiple baselines.    In either case, the  beam combiner takes light from each of the input waveguides and splits it equally into  two output fibres, such that each output fibre contains a fixed fraction of the light from each telescope, which interfere.  By varying the optical path length in one arm, and recording the output intensity, fringes can be recorded, and from these the visibility can be computed.

A significant  advantage of using waveguides to accomplish the beam combination is the spatial filtering provided by single-mode waveguides, which has led to an order of magnitude improvement in the ability to recover visibility fringes\cite{coud98}.  In general, the atmospheric turbulence above each telescope will destroy any coherence between the two telescopes, and recovering visibilities will have to rely on statistical calibration methods.
However, all spatial information is lost when coupling light into a single-mode fibre; there is only one mode, and the changing seeing will not change the coherence of light coupled via single-mode waveguides.
The coupling efficiency into the SMF will change with as the seeing changes, but this can be calibrated with a Y-splitter (section~\ref{sec:combine})to record the photometry.

Today,the state-of-the-art in optical interferometry is found in the GRAVITY instrument\cite{grav17} on the VLT.  This combines the light from $4 \times 8$~m telescopes using a phosphor-doped silica on a silicon wafer manufactured by the Laboratoire d'\'{E}lectronique des Technologies de l'Information (CEA/LETI) using plasma-enhanced chemical vapor deposition\cite{joc14}.

Following adaptive optics correction, performed individually for each telescope, GRAVITY injects the light from two separate objects into optical fibres.
These fibres (two per telescope) feed into the integrated optics beam combiners.  These beam combiners incorporate phase shifting beam splitters in the waveguides, such that each beam combiner can sample the interference at four different phases, in  technique known as ABCD beam combination.  

The 24 outputs from each combiner each feed into a spectrometer.  One of the spectrometers, also contains a camera which is read out at kHz frequncy to measure the fringes in real time and correct for the effects of the atmosphere.
 The second spectrometer is then used for long science exposures (hundreds of seconds), which would otherwise be restricted to a few milliseconds.   The two feeds can also be used for extremely accurate (microarcsecond) astrometry between pairs of objects.  
 
 GRAVITY has tested the equivalence principle of general relativity in hitherto untested regimes through observations of stellar orbits around the super massive black hole at the centre of our Galaxy\cite{grav19}, and has made the most detailed observations of material orbiting close to a black hole, measuring  velocities of $\approx 30$\% the speed of light\cite{grav18}.  These unprecedented observations contributed to 2020 Nobel Prize, and  are a testament to the power of photonics in astronomical interferometry.

\subsection{Quantum optical interferometry}
\label{sec:quantuminter}

\begin{figure}[t]

    \centering
    \includegraphics[scale=0.35]{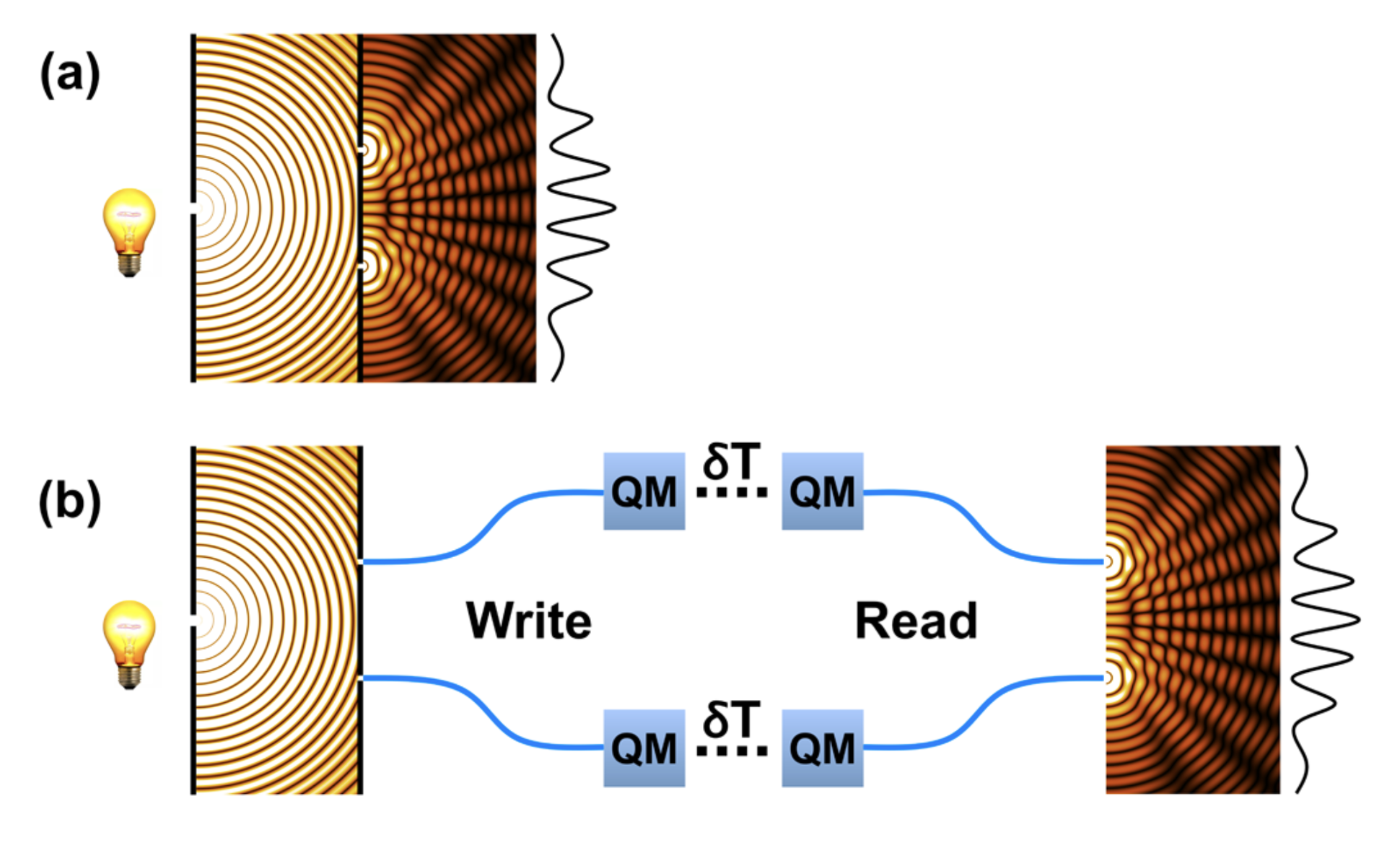}
\caption{(a) Classical Young's slit experiment. A light source illuminates a narrow slit. The quasi-coherent slit beam propagates towards two narrow slits {\bf 1} and {\bf 2}. The beams from these slits interfere and generate a fringe pattern at a distant screen. Remarkably, if we replace the screen with a detector and take a long exposure, similar fringes can be seen if the illuminating source emits only one photon at a time.
(b) Sketch of a quantum telescope interferometer taken from Bland-Hawthorn et al 2021\cite{bland21}. This is a more practical and less expensive implementation of a quantum network by using quantum memories to store light. By analogy with the Event Horizon Telescope, these can be brought together at a later time and read out to form a signal. (A) The conventional Young’s double slit experiment illustrating how a coherent source generates an interferogram at a distant screen, assuming the slits are diffraction limited. The line trace is the amplitude of the electric field. (B) To illustrate experiment is now split into two parts. Two single-mode fibres act as the double slit and direct photonic states to be written into independent quantum memories (QMs). The ellipses represent time passing ($\delta t$), after which the photonic data stored in the QMs are read out into single-mode fibres and their well-aligned outputs are allowed to interfere. The interferogram is recovered for which every detected photon arises from a superposition of information stored in both QMs.  }
    \label{f:QMslits}
\end{figure}

The ideas on optical interferometry described in section~\ref{sec:interferometer} can be extended to extremely long baselines using the ideas from quantum mechanics and quantum computing described in section~\ref{sec:qm}.  The fundamental concept here is 
the
quantum mechanical interpretation of Young's slits.
In Fig.~\ref{f:QMslits}(a), an interference effect is visible after many events have been detected by the apparatus, even if those events are detected from individual particles. If $\langle x\vert$ describes the position $x$ on the screen, and $\vert \psi\rangle$ describes the energy and momentum of the radiation via the pdf (wave function), we can write down an expression for the Young's slit experiment:
\begin{equation}
    \psi(x) = \langle x\vert\psi\rangle = \langle x\vert {\bf 1}\rangle \langle {\bf 1}\vert\psi\rangle  +\langle x\vert {\bf 2}\rangle \langle {\bf 2}\vert\psi\rangle 
\end{equation}
where $\langle {\bf j}\vert \psi\rangle$ is the probability (amplitude) that the particle travelled through slit {\bf j}, and $\langle x\vert {\bf j}\rangle$ is the probability the particles reaches $x$ after leaving slit {\bf j} (Fig.~\ref{f:QMslits}(a)). This equation describes a superposition; in essence, the photonic state interferes with itself at the detector. 

So how do we conduct a Young's slit experiment with single photon events for widely spaced receivers? Carrying a coherent photonic state over an optical fibre is impossible for long distances, and difficult even for expensive polarising fibres over short distances. Here we present two new ideas borrowing from developments in quantum computing.  First, we look at a quantum network, in which the knowledge of the state of a particle may be transmitted to a distant site via entanglement.  Secondly, we look at using quantum memories to store the state of a particle (without measurement), prior to interference.  

\subsubsection{Quantum networks}

First let us look at quantum networks.  How is it possible to send the state of a particle to a distant site, without measuring it?

\begin{figure}
    \centering
    \includegraphics[scale=0.4]{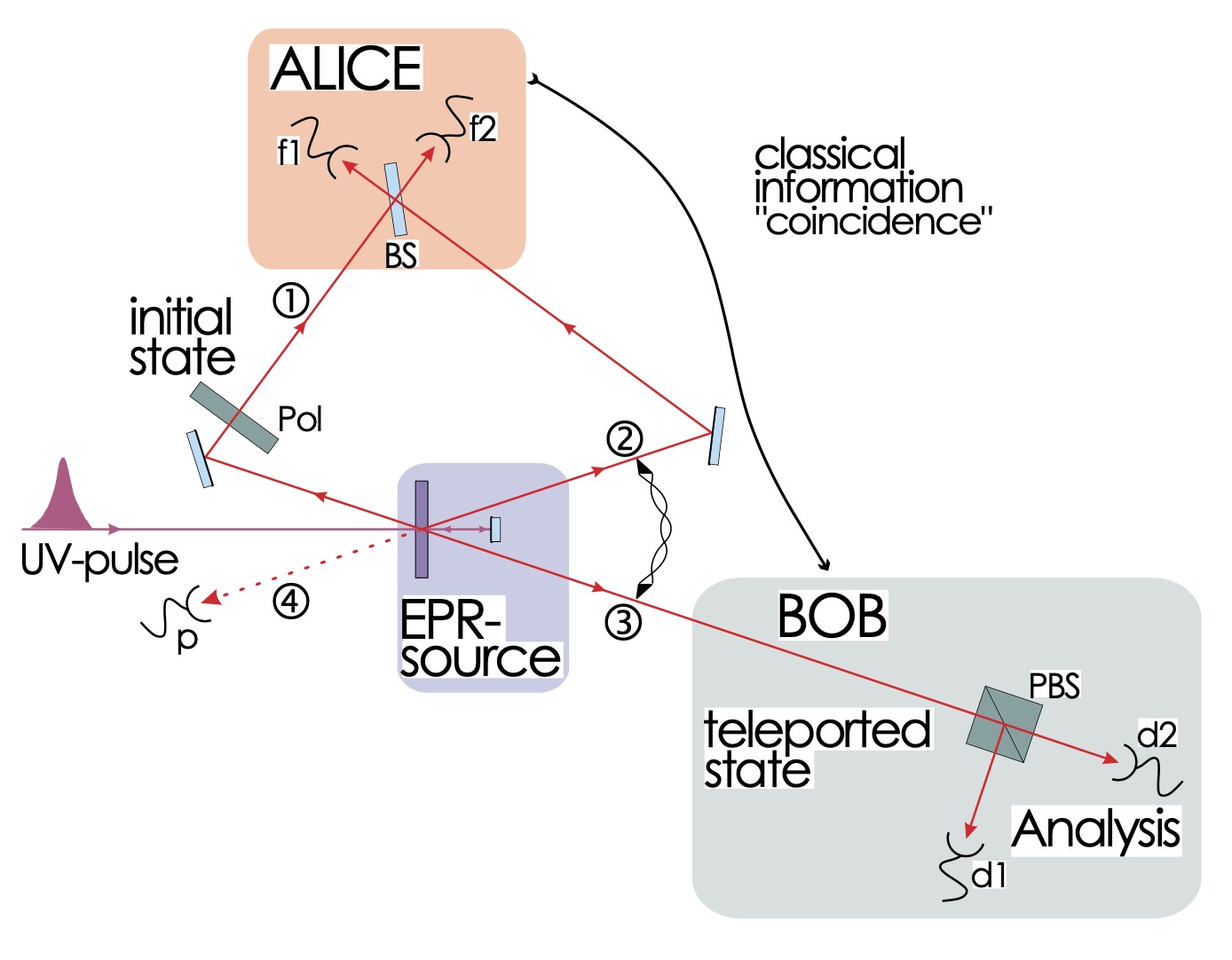}
    \caption{Sketch of a quantum teleportation circuit\cite{wei01} that simplifies the complex layout of actual circuits\cite{ada99,lle20}. An input 405 nm photon generates two entangled photons; these travel along quantum-preserving channels \circled{2} and \circled{3}, one per channel, to Alice and Bob. The entangled photons are each superposed with other states on arrival. Alice communicates openly with Bob about what logic gate to apply in order to understand her teleported information.}
    \label{f:weihs}
\end{figure}

Following the language of quantum technology, Alice wants to communicate the state of a particle to  Bob, who is located at a distant site, see Fig.~\ref{f:weihs}, without measuring that state.

The sequence of steps are
\begin{itemize}
    \item Alice initializes her qubit $\vert\psi\rangle_s$ to the informative state she wishes to communicate.
    \item A Bell state (two-qubit state) is created by an entanglement source; one qubit $\vert\psi\rangle_A$ is sent to Alice, the other $\vert\psi\rangle_B$ to Bob.
    \item Alice converts her two qubits from the Bell basis to a new basis.
    \item Alice measures her two qubits.
    \item Alice tells Bob how to convert his qubit (on an open line), i.e. what gate to use.
\end{itemize}
This all sounds rather opaque until we break it down. Let us assume a starting three-qubit state of 
\begin{equation}
     \vert \psi\rangle_s\vert \psi\rangle_A\vert \psi\rangle_B = \vert \psi\rangle
     _s\vert 0\rangle_A\vert 0\rangle_B
\end{equation}
where we have yet to define the first qubit. After the A and B qubits become entangled, the combined state (known as a Bell state) is
\begin{equation}
    \vert\Psi\rangle = \frac{1}{\sqrt{2}}( \vert \psi\rangle
     _s\vert 0\rangle_A\vert 0\rangle_B + \vert \psi\rangle
     _s\vert 1\rangle_A\vert 1\rangle_B)
     \label{e:tele1}
\end{equation}
Let Alice's initial state now be $\vert\psi\rangle_s = \alpha\vert 0\rangle + \beta\vert 1\rangle$, with $\alpha$ and $\beta$ as the usual complex amplitudes. Substituting this into Eq.~\ref{e:tele1}, and dropping subscripts, 
\begin{equation}
     \vert\Psi\rangle = \frac{1}{\sqrt{2}}(\alpha\vert 000\rangle + \beta\vert 100\rangle + \alpha\vert 011\rangle + \beta\vert 111\rangle)
\end{equation}
Our optical circuit can now apply a $\boxed{\mathbf{CNOT}}$ (equation~\ref{eqn:cnot}) operation to the first two qubits, such that 
\begin{equation}
     \vert\Psi\rangle = \frac{1}{\sqrt{2}}(\alpha\vert 000\rangle + \beta\vert 1\stackrel{\downarrow}{1}0\rangle + \alpha\vert 011\rangle + \beta\vert 1\stackrel{\downarrow}{0}1\rangle)
\end{equation}
where the flipped bits are indicated. This is
followed by a Hadamard gate (equation~\ref{eqn:hadamard}) on the first qubit, so that
\begin{equation}
     \vert\Psi\rangle = \frac{1}{2}(\alpha\vert 000\rangle + \alpha\vert 100\rangle + \beta\vert 010\rangle - \beta\vert 110\rangle + \alpha\vert 011\rangle + \alpha\vert 111\rangle + \beta\vert 001\rangle - \beta\vert 101\rangle).
\end{equation}
The beauty of the method is revealed when we rearrange the above equation:
\begin{equation}
\vert\Psi\rangle = \frac{1}{2}\mathbf{\vert 00}\rangle(\alpha\vert 0\rangle + \beta\vert 1\rangle) + \frac{1}{2}\mathbf{\vert 01}\rangle(\beta\vert 0\rangle + \alpha\vert 1\rangle) 
+ \frac{1}{2}\mathbf{\vert 10}\rangle(\alpha\vert 0\rangle - \beta\vert 1\rangle) 
+ \frac{1}{2}\mathbf{\vert 11}\rangle(-\beta\vert 0\rangle + \alpha\vert 1\rangle)
\end{equation}
where the Bell basis is highlighted as bold text. 
All four variants of Alice's prepared state are equally possible outcomes when Alice measures her two-qubit state. Let's say she measures  $\vert 00\rangle$, then she knows Bob received her prepared state $\vert\psi\rangle_s = \alpha\vert 0\rangle + \beta\vert 1\rangle$ - success! But if she measures $\vert 01\rangle$, she knows his qubit state $\vert\psi_B\rangle$ needs to be corrected with an X-gate (see section~\ref{sec:entangle}), and so on...
\begin{eqnarray}
    \vert \psi\rangle_s &=& \mathbf{I}\vert \psi_B\rangle\;\;\;\;\;\;\;\;\;\;\: \rm (Alice\; measures\; \vert 00\rangle)\\
    \vert \psi\rangle_s &=& \mathbf{X}\vert \psi_B\rangle\;\;\;\;\;\;\;\;\; \rm (Alice\; measures\; \vert 01\rangle)\\
    \vert \psi\rangle_s &=& \mathbf{Z}\vert \psi_B\rangle\;\;\;\;\;\;\;\;\;\:\rm (Alice\; measures\; \vert 10\rangle)\\
    \vert \psi\rangle_s &=& \mathbf{ZX}\vert \psi_B\rangle\;\;\;\;\;\;\; \rm (Alice\; measures\; \vert 11\rangle)
\end{eqnarray}
Alice communicates the required operator on an open channel to Bob to ensure success for all eventualities. 

\paragraph{Quantum networks for interferometry}
\label{sec:qnetint}

This principle can be extended to establish
the use of quantum networks that ensure multiple nodes are able to communicate with coherent states, i.e.\ qubits  (Gottesman et al 2012)\cite{got12}.


The principle behind this proposal is that we can determine the mutual coherence between $N$ telescopes by quantum entanglement.  Just as single photon events generate fringes when summed over many events, in a two-telescope configuration, interference is only possible if a single photon is detected at either telescope, where we remain ignorant of which one. If photons from the star arrive at both telescopes simultaneously, the uncertainty disappears and this does not contribute signal to the fringe visibility.

\begin{figure}
    \centering
    \includegraphics[scale=0.3]{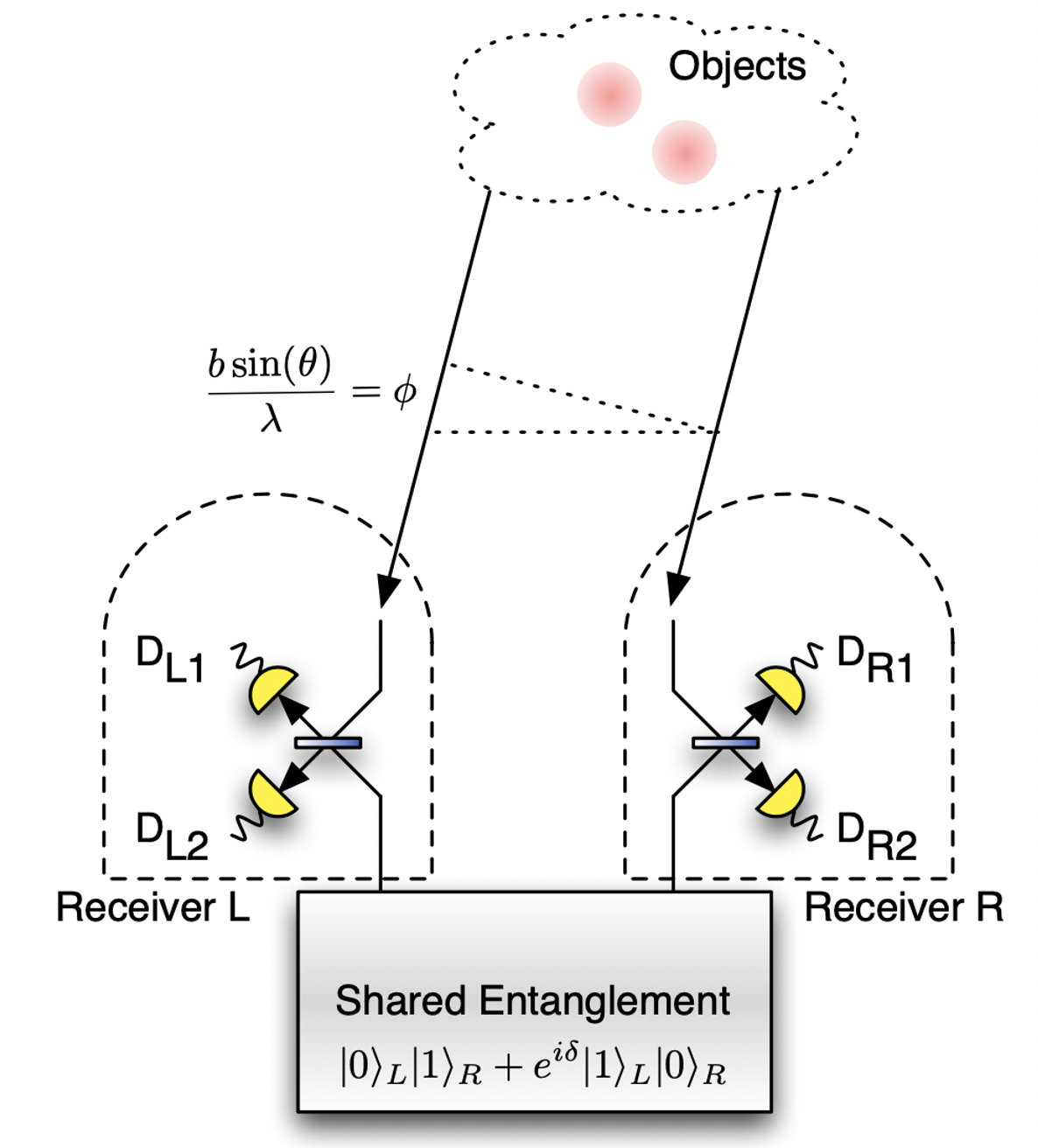}
    \caption{Sketch of a quantum telescope interferometer taken from Gottesman et al 2012. 
    The entangled photon pair are sent in different directions to the separate telescopes. The beam splitter ``mixes'' the incoming science photon with one of the entangled pair to achieve a superposition across both telescopes.}
    \label{f:gott1}
\end{figure}

So in order to achieve interference fringes over a long baseline while remaining ``in the dark,'' we turn to quantum networks.
This mutual coherence is maintained by a source of entangled photons, where one of the pair is sent to each telescope (see Fig.~\ref{f:gott1}). One of the pair can be delayed in time ($\delta t$) to `simulate' the effect of varying a baseline. In the language of quantum mechanics, these serve as {\it non-local} oscillators. 
Networks can also assist with accurate timing between distant nodes using quantum clock synchronization protocols. With sufficiently fine time sampling, even very low rates of entangled photons against a strong noisy background generate strong signals in the correlated data.

\begin{figure}
    \centering
    \includegraphics[scale=0.425]{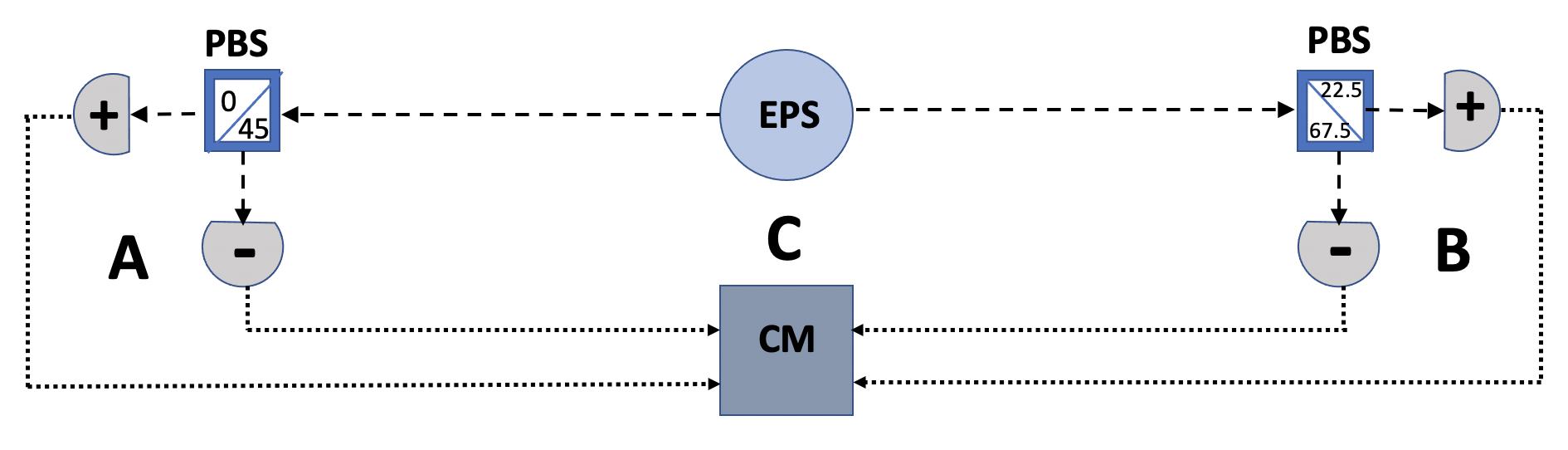}
    \caption{The key to ensuring that two distant sites A and B are entangled is to carry out a Bell state (or inequality) test. An entangled photon source (EPS) from a third party site C produces pairs of photons where one each is sent to two distant sites A and B. A polarising beamsplitter (PBS) is arranged with two possible polarisations as shown. In the absence of noise, only one of the two detectors at A indicated by $\pm$, and one at B, is expected to register a photon. All four time series of detector clicks are  recorded by a coincidence monitor (CM) at C. After conducting four independent experiments with the PBS orientations (see text), a simple statistical test determines if entanglement has been achieved between A and B.}
    \label{f:chsh}
\end{figure}
Fig.~\ref{f:chsh} illustrates
a key aspect of Gottesman's idea, which builds on a fundamental principle in quantum networks. In order to pass delicate quantum information over a network, each node must be in a superposition (mutually coherent) with the other node. In the absence of noise, the four detector outputs allow for only four realizations ($++, --, +-, -+$).
These are monitored by the third party as shown, ideally with picosecond timing accuracy. For any combination of two polarisations, the corresponding accumulated counts are $N_{\scriptscriptstyle ++}$,\; $N_{\scriptscriptstyle --}$,\; $N_{\scriptscriptstyle -+}$,\; $N_{\scriptscriptstyle +-}$.
Four experiments $E_i$ are carried out for all four intercombinations of the polarising beam splitters (PBS).
The PBS orientations are set to $E_1(0,22.5),\; E_2(0,67.5),\; E_3(45,22.5),\; E_4(45,67.5)$. For each of these, we record
\begin{equation}
    E_i = \frac{(N_{\scriptscriptstyle ++} + N_{\scriptscriptstyle --}) - (N_{\scriptscriptstyle -+} + N_{\scriptscriptstyle +-})}{(N_{\scriptscriptstyle ++} + N_{\scriptscriptstyle --}) + (N_{\scriptscriptstyle -+} + N_{\scriptscriptstyle +-})}
\end{equation}
The Bell state test for entanglement is then
\begin{equation}
S = E_1 - E_2 + E_3 + E_4 > 2.
\end{equation}
In a properly constructed experiment, this condition has been shown to be true in many experiments, i.e.\ that quantum non-locality is a real phenomenon and entangled particles maintain their superposition at any physical separation.
If $N_i$ are the data for each $E_i$, we can derive relations for $\partial S/\partial N_i$ for different sample sizes to determine an error $\sigma_S$ for $S$.
Thus, if shot noise dominates,
\begin{eqnarray}
    \sigma_S^2 &=& \sum_i \left(\sigma_{N_i} \frac{\partial S}{\partial N_i} \right)^2 \\
    &=& \sum_i N_i \left(\frac{\partial S}{\partial N_i} \right)^2
\end{eqnarray}
where $i$ can be varied over as many polarisation states as required\cite{deh02}. The four angles quoted above give a maximal Bell state in modern lab experiments of $S \approx 2.5$ with of order 0.1\% uncertainty\cite{ves22}, where $S = 2\sqrt{2} \approx 2.8$ is the theoretical maximum.

Through the concept of {\it entanglement swapping}, it even becomes possible to form a daisy chain with the A-C-B layout, in the sense of A-C$_1$-B-C$_2$-D-C$_3$-E... such that A, B, D, E, etc. are all in a coherent superposition. This is basis for the quantum repeater technology that maintains the quantum network with the addition of quantum memories (see below) at each node to store the entangled particles. We are still a long way from a true quantum internet, but rudimentary networks exist. In the next section, we present our alternative approach to achieving an astronomical quantum network.

In Fig.~\ref{f:gott1}, let us imagine that a science photon has arrived on the left side. Within a coherence time, it enters the beam splitter with the entangled photon, and forms a superposition with the other member of the pair (witness photon) at the right side, which, for the precisely known delay time, also arrives within a coherence time. The quantum mechanical superposition can be written as
\begin{equation}
    \vert \psi \rangle =\vert 0\rangle_L \vert 1\rangle_R + e^{i\delta} \vert 1\rangle_L \vert 0\rangle_R
\end{equation}
where $\delta$ is the optical path delay.
We collect data at both telescopes to ensure this is working correctly with full knowledge of the delay $\delta t$ that was imposed. The signal is not measured from the number of events, but from the correlations across the four detector clicks. The beam splitters are essentially ``Bell inequality'' tests that ensure the entanglement.
The beauty of the method is that the detector correlations build up a fringe pattern as a function of $\delta\; (=c\;\delta t)$ if and when a science photon arrives at one, and only one (either will do), telescope. Of course, there will be sources of noise that do not build constructively, but lab experiments show a strong coherence detection even at low signal-to-noise ratios.

\subsubsection{Quantum memories}

Now we look at a second method for quantum optical interferometry, viz.\ using quantum memories to store the states of photons received by the telescopes prior to interference.

In our earlier mention of quantum repeater protocols (section~\ref{sec:qnetint}), we mentioned the concept of a quantum memory (QM) for storing entangled photons at each node. These are also central to our new proposal for a quantum telescope interferometer, discussed below, and thus deserve a brief discussion.
Stopping a photon in its tracks presents a difficult problem. Indeed, in astronomy, all successful methods to date are based on trapping a photon (strictly, an energy packet) in an optical fibre coil or a high-$Q$ optical cavity. The best telecom fibres have 0.14 dB/km loss such that $>$95\% of photons are lost after propagating 100 km (storage time $\sim$ 0.5 ms). The best optical cavities allow up to a million or more internal reflections before photons are absorbed or scattered. In both instances, the trapping time is short ($\ll 1$ sec) before the photon is eventually lost to the system. A much better option is to couple light quanta with matter particles. 

Quantum memories rely on advanced ideas in cavity quantum electrodynamics\footnote{This field was recognized by the 2012 Nobel Prize in Physics awarded to Serge Haroche and David Wineland. For a review, we refer the reader to \cite{wal06}.} and so we supply only a basic introduction here. 
So what kind of light-matter interaction allows for a photon's quantum state to be transferred to an atomic state? Light-matter interactions are typically very weak: to achieve a high probability of the photonic information being written onto a single atom, we first consider an energy packet strongly interacting with an atom embedded within a high-$Q$ optical cavity. The atom must then be safeguarded against its noisy environment during storage because the fragile quantum state can be easily disrupted. 
Specht et al.\cite{spe11} realized a read-write QM using a single Rb atom trapped within an optical cavity. 

But the first demonstration of a QM for light with an efficiency greater than 50\% was enabled by mapping photons onto the collective state of a large ensemble of atoms `trapped' in a solid-state crystal~\cite{hed10}. Here the strong light-matter interaction was achieved by using $10^{13}$ atoms, overcoming the very weak interaction at a single atom level. In Bland-Hawthorn et al.\cite{bland21}, we discuss a proposal that combines elements of these two approaches - coupling ensembles of atoms in the solid-state to high-$Q$ photonic cavities - that allows for the storage of large numbers of photons and for the memory to be transported, while demonstrating the potential for storage times that are interesting for practical applications.



Consider the two experiments illustrated in Fig.~\ref{f:QMslits}. The upper apparatus in (A) is the conventional Young's double-slit experiment where a diffraction limited source (e.g. laser) or any source illuminating a diffraction-limited slit illuminates two slits at the next screen. 
In the limit of low illumination (the single particle regime), if an attempt is made to measure which slit the particle passed through, the interference pattern is lost, in agreement with Heisenberg's uncertainty principle. 

We now propose an equivalent experiment in (B) that has profound implications for astronomical interferometry and related fields in physics. The two slits are replaced with independent storage systems (QMs) that are able to preserve input photonic quantum states (the phase and amplitude of light as a function of time). At a later time, if the QMs can be read out in synchrony to faithfully recall the stored photons, the original interference pattern is recovered. The write and read operations in Fig.~\ref{f:QMslits} can be executed {\it without measurement} allowing the photonic quantum state to be preserved. 



\section{Summary}
\label{sec:conc}

Today, we stand at the dawn of a new era of extremely large ($25-40$~m)  telescopes. It is conceivable that this scaling-up will continue
until there is a breakthrough in the underlying technology or a fundamental limit is reached.   These telescopes will have such enormous light-gathering potential that the time has come to consider new ways of doing astronomy.

We have introduced photonic functions that provide the foundation of astrophotonics and reviewed some key astrophotonic instruments currently in development.
The breadth of development is already wide and significant, and is expanding.  Astrophotonics is in a period of transition, as many instruments are progressing from concepts to prototypes to full facility class instruments, and the pace of development is likely to increase as a greater number of photonic technologies are adapted for astronomy. In this respect, like many science fields (e.g. medical photonics, biophotonics), astrophotonics will benefit from further progress in detector development, especially photodetector integration onto photonic waveguides to aid progress.

Some astrophotonic instruments have already had a transformative impact on astrophysics, in particular, the gains resulting from photonic interferometric beam combiners (section~\ref{sec:interferometer}), and several others are on the cusp of making a similar impact, e.g.\ OH suppression (section~\ref{sec:ohsupp}).
Here, we have also looked to the future, with the development of quantum astrophotonic instruments.  The possibility of recording the quantum state of individual photons, or communicating this over quantum networks, gives rise to the possibility for extremely long baseline, optical interferometry.

Astrophotonics can play a key role in this development, offering radical new ways to manipulate  light, far beyond the traditional iterations and scaling-up of previous instruments.
Indeed, astrophotonics is especially apposite to the ELT era. Adaptive optics and near-diffraction limited performance are designed into 
the telescopes at the outset, with the effect that coupling into astrophotonic devices will be much simpler and more efficient.  Moreover, the ease of replication and the modularity of photonic components allows a new approach to instrumentation that can end the perpetual increase of instrument size, and telescopes get bigger.  That is to say, ELTs will make astrophotonics easier, and in return astrophotonics will make ELTs more productive and powerful.

\section{Acknowledgments}
We are grateful to many colleagues at the Sydney Astrophotonic Instrumentation Labs at the University of Sydney, Australian Astronomical Optics, and at Australian Astronomical Optics at Macquarie University. We are particularly indebted to our co-author Sergio Leon-Saval for allowing us to use points of discussion in our recent textbook ``Principles of Astrophotonics.'' JBH is indebted to John Bartholomew and Matt Sellars for our ongoing collaborative developments on quantum telescopes.


\bibliographystyle{tfnlm}
\bibliography{ps}

\end{document}